\documentclass[12pt]{article}
\usepackage{axodraw,epsfig}
\begin{document}

\def\bef{\begin{figure}}
\def\eef{\end{figure}}
\newcommand{\ans}{ansatz }
\newcommand{\be}[1]{\begin{equation}\label{#1}}
\newcommand{\beq}{\begin{equation}}
\newcommand{\ee}{\end{equation}}
\newcommand{\beqn}[1]{\begin{eqnarray}\label{#1}}
\newcommand{\eeqn}{\end{eqnarray}}
\newcommand{\bd}{\begin{displaymath}}
\newcommand{\ed}{\end{displaymath}}
\newcommand{\mat}[4]{\left(\begin{array}{cc}{#1}&{#2}\\{#3}&{#4}
\end{array}\right)}
\newcommand{\matr}[9]{\left(\begin{array}{ccc}{#1}&{#2}&{#3}\\
{#4}&{#5}&{#6}\\{#7}&{#8}&{#9}\end{array}\right)}
\newcommand{\matrr}[6]{\left(\begin{array}{cc}{#1}&{#2}\\
{#3}&{#4}\\{#5}&{#6}\end{array}\right)}
\def\lsim{\raise0.3ex\hbox{$\;<$\kern-0.75em\raise-1.1ex
\hbox{$\sim\;$}}}
\def\gsim{\raise0.3ex\hbox{$\;>$\kern-0.75em\raise-1.1ex
\hbox{$\sim\;$}}}
\def\abs#1{\left| #1\right|}
\def\simlt{\mathrel{\lower2.5pt\vbox{\lineskip=0pt\baselineskip=0pt
           \hbox{$<$}\hbox{$\sim$}}}}
\def\simgt{\mathrel{\lower2.5pt\vbox{\lineskip=0pt\baselineskip=0pt
           \hbox{$>$}\hbox{$\sim$}}}}
\def\unity{{\hbox{1\kern-.8mm l}}}
\def\epr{E^\prime}
\newcommand{\al}{\alpha}
\def\16p{16\pi^2}
\newcommand{\eps}{\varepsilon}
\newcommand{\epsr}{\varepsilon_{ R}}
\newcommand{\epsl}{\varepsilon_{ L}}
\newcommand{\epsrs}{\varepsilon_{s R}}
\newcommand{\epsls}{\varepsilon_{s L}}
\def\ep{\epsilon}
\def\ga{\gamma}
\def\Ga{\Gamma}
\def\om{\omega}
\def\OM{\Omega}
\def\la{\lambda}
\def\La{\Lambda}
\def\al{\alpha}
\def\bY{{\bf Y}}
\def\bA{{\bf A}}
\def\bU{{\bf U}}
\def\bm{{\bf m}}
\def\bM{{\bf M}}
\newcommand{\ov}{\overline}
\renewcommand{\to}{\rightarrow}
\renewcommand{\vec}[1]{\mbox{\boldmath$#1$}}
\def\tm{{\widetilde{m}}}
\def\tl{{\tilde{L}}}
\def\te{{\tilde{e^c}}}
\def\tel{{\tilde{e}}}
\def\td{{\tilde{d^c}}}
\def\tq{{\tilde{Q}}}
\def\tu{{\tilde{u^c}}}
\def\mcirc{{\stackrel{o}{m}}}
\def\dem{\delta m^2} 
\def\sint{\sin^2 2\theta} 
\def\tant{\tan 2\theta} 
\def\tanL{\tan 2\theta^L}
\def\tanR{\tan 2\theta^R}
\newcommand{\tanb}{\tan\beta}
\def\dfrac#1#2{{\displaystyle\frac{#1}{#2}}}

%

\begin{titlepage}

\begin{flushright}

DFPD-02/TH/15 \\ 

\today
\end{flushright}

\vspace{2.0cm}

\begin{center}

{\Large \bf 
Supersymmetric Seesaw without Singlet Neutrinos: \\
\vspace{0.4cm}
Neutrino Masses and 
Lepton-Flavour Violation}

\vspace{0.7cm}

{\large 
Anna Rossi
}
\vspace{5mm}

{\it  Dipartimento di Fisica `G.~Galilei',  
Universit\`a di Padova and \\
INFN, Sezione di Padova, Via Marzolo 8, I-35131 Padua, Italy\\
{\rm E-mail address: arossi@pd.infn.it } 
}
\end{center}

\vspace{10mm}

\begin{abstract}
\noindent
We consider the supersymmetric seesaw mechanism 
induced by the exchange of heavy  
$SU(2)_W$ triplet states, rather than `right-handed' neutrino singlets,
to generate neutrino masses. 
We show that 
in this scenario the neutrino flavour structure  tested at low-energy 
in the atmospheric and solar   neutrino experiments is 
directly inherited from the 
neutrino Yukawa couplings to the triplets. This allows us to predict 
the ratio of the  $\tau \to \mu \ga$ (or $\tau \to e \ga$) and 
$\mu\to  e \ga$ decay rates in terms of the low-energy  neutrino 
parameters. Moreover, once the model is embedded in a grand unified model, 
quark-flavour violation can be linked to lepton-flavour violation.

\end{abstract}
\end{titlepage}

\setcounter{footnote}{0}

\section{Introduction}
Nowadays we have one more important piece of information 
concerning the flavour structure of the Standard Model:  flavour 
violation is also present  in the lepton sector. 
So far this has only shown up in the atmospheric and 
solar neutrino 
experiments \cite{atm,solar}. The anomalies observed in these experiments 
can be interpreted in terms of neutrino oscillations 
which are the result of non-vanishing neutrino masses and mixing angles 
\cite{P,matter}.
In the framework of the Standard Model (or of its supersymmetrized 
version) the latter properties effectively arise from 
the following  lepton-number ($L$)  violating $d=5$ operator \cite{d5}:
 \be{d5}
\frac{1}{2M_L}  \bY_\nu^{i j} (L_i H_2) (L_j H_2) ,~~~~~~~ \bY_\nu =\bY^T_\nu
\ee
where $i,j =1,2,3$ are family indices, 
$M_L$ is the energy scale where $L$ is broken, 
$L_i$ are  the $SU(2)_W$ lepton doublets 
and $H_2$ is the Higgs doublet with hypercharge $Y=1/2$.
Upon breaking $SU(2)_W\times U(1)_Y$ by the vacuum expectation value 
of the Higgs field, $\langle H_2\rangle = v_2= v \sin\beta$ 
($v=174 {\rm GeV}$), the 
operator (\ref{d5}) induces Majorana masses for the neutrinos,
\be{nu-mass}
{\bf m}^{ij} _\nu = \frac{v_2^2}{M_L} \bY_\nu^{i j} .
\ee 
Therefore, from here we can easily understand the origin of the tiny neutrino 
masses, as they may be suppressed by some large scale $M_L$. 
Taking for instance a neutrino mass $m_\nu \sim (10^{-1}-10^{-2})~{\rm eV}$  
as indicated by the atmospheric neutrino data (and assuming 
a hierarchical neutrino spectrum) \cite{analysis}, 
the magnitude of $M_L$ can be inferred, 
\be{Ml}
Y^{-1}_\nu M_L \sim 10^{15} ~{\rm GeV} .
\ee
In the basis in which  
the charged-lepton Yukawa matrix $\bY_e$ 
is diagonal, all the lepton-flavour violation is contained 
in the coupling matrix $\bY_\nu$, i.e. in the neutrino mass matrix:
\be{mix}
 \bm_\nu  
= {\bf U}^\star\bm^D_\nu {\bf U}^\dagger , ~~~~~~~~~
\bm^D_\nu = {\rm diag}(m_1, m_2, m_3)
\ee
where $m_1, m_2, m_3$ are the neutrino mass eigenvalues, 
and the unitary matrix ${\bf U}$ is the   lepton mixing matrix 
that  
appears in the charged lepton current $\bar{\ell} \ga^\mu (1 -\ga_5)
\bU~ \nu$ and is  responsible for neutrino oscillations.    

In principle lepton-flavour violation could also be tested in other processes, 
such as $\mu\to e\ga$, $\tau\to \mu \ga$ and $\tau\to e \ga$. 
The present experimental limits on these decays are \cite{limit}:
\beqn{limit}
BR(\mu \to e\ga) &<& 1.2 \times 10^{-11} , \\ \nonumber
BR(\tau \to \mu \ga) &<& 1.1 \times 10^{-6} , \\ \nonumber
BR(\tau \to e \ga) &<& 2.7 \times 10^{-6} , 
\eeqn
which could be significantly improved, the first down to $10^{-14}$ 
\cite{mue} 
and the second to $10^{-9}$ \cite{taumu}.
In the Standard Model 
it is  very hard 
to obtain interesting results because the 
decay amplitudes are strongly GIM-suppressed 
by the tiny neutrino masses, 
$m_i \ll M_W$ \cite{mueg}.  
On the other hand, 
in the framework of the minimal supersymmetric extension of the 
Standard Model (MSSM), those  processes can be enhanced 
through the one-loop exchange of  superpartners, if the masses of
the latter are not too heavy and do not conserve flavour \cite{fcmssm}. 
Concerning the latter property, 
we could expect  that 
the underlying flavour theory dictates both the flavour structure 
of  the standard fermion mass matrices 
and that of the corresponding supersymmetric 
scalar partners (see e.g. \cite{FT}).
Even if sfermion masses are universal  
(i.e.  flavour-blind) 
at high energy, as in 
minimal supergravity or gauge mediated supersymmetry breaking scenarios, 
nevertheless flavour conservation can be broken  
 in the sfermion masses  
by radiative effects due to  flavour-violating 
Yukawa couplings \cite{old,dec}.
In particular, the interactions that generate the operator 
(\ref{d5}) also induce 
lepton-flavour violation (LFV) in the slepton mass matrices 
by renormalization effects. 
A well-known and investigated 
example is that of the standard seesaw mechanism \cite{ssn}
in which LFV is induced radiatively by  
Yukawa couplings of 
the $SU(2)_W$-doublet neutrinos with singlet neutrinos $N$ (often 
referred  to as right-handed neutrinos) \cite{old, new}. 
Here, we would like to discuss another example of seesaw scenario 
in which the $d=5$ operator (\ref{d5}) is obtained by 
exchanging  heavy $SU(2)_W$ triplets with non-zero hypercharge. 
We recall that models with scalar triplets to 
generate Majorana neutrino masses have been considered 
in the literature for a long time, though they have received 
less attention than the standard seesaw scenario.
For example,  a model with spontaneous 
$L$-breaking was proposed in \cite{GR} 
and later on extended   \cite{SV}.
A triplet-exchange seesaw realization with explicit $L$-breaking 
was  also introduced \cite{TS1},  
in a non-supersymmetric framework\footnote{
Recently, this scenario was studied for leptogenesis \cite{TS2}. 
An alternative seesaw mechanism, obtained by exchanging 
heavy $SU(2)_W$ triplets with zero hypercharge, was discussed in 
\cite{ma}.}.  
A supersymmetric version of the 
latter scenario was recently introduced to have baryogenesis 
through leptogenesis \cite{TS3}. 
In this work we shall further elaborate the supersymmetric triplet seesaw 
scenario.
In particular, we shall 
discuss how LFV is radiatively induced in the slepton masses and 
show that this scenario is potentially more predictive 
than the $N$-induced seesaw one.

This paper is organised as follows. In Sect. 2 we 
review both 
the standard see-saw mechanism and that in which triplet states 
are exchanged. This presentation will help in showing the 
differences between the two scenarios and outline 
the one-to-one correspondence between the neutrino parameters and 
LFV in the soft-breaking parameters that characterizes the triplet 
seesaw scenario.
In Sect. 3 the triplet seesaw is embedded in $SU(5)$ context. 
In Sect. 4 we show the general pattern of $\bY_\nu$ as derived from 
the low-energy neutrino data.
In Sect. 5 we discuss 
the flavour-violation induced in the soft-breaking terms 
by radiative effects in the energy range above the triplet 
mass threshold. 
The renormalization group equations (RGEs) relevant 
to our study are confined in Appendix.
In Sect. 6, after giving the qualitative  
behaviour of the $\ell_i \to \ell_j  \gamma$ 
branching ratios,  we also discuss some numerical examples. 
We give our conclusions in Sect. 7.

\section{Singlet versus Triplet seesaw}
First, we briefly review the standard seesaw mechanism 
in which singlet states $N$ are exchanged \cite{ssn}.
The relevant superpotential terms at the scale where 
lepton number is broken are:
\be{L-N}
\bY^{ij}_{N}N_i L_j H_2 + \frac{1}{2} {\bf M}^{ i j }_N N_i N_j  ,
\ee
where $i,j =1,2,3$ are family indices, 
$\bY_N$ is an arbitrary 3$\times$ 3  matrix of 
Dirac-like Yukawa couplings, while $\bM_N$ 
is a 3$\times$ 3 symmetric mass matrix describing Majorana masses 
for the singlets $N$.    
If the $N$ states are  assigned  lepton number $L=-1$, 
the second term in (\ref{L-N}) 
explicitly breaks $L$. 
After decoupling the heavy states $N$ at the (overall) scale $M_L$, 
the lepton-number violating $d=5$ operator (\ref{d5}) is generated,
where $M^{-1}_L \bY_\nu$ is identified as follows:
\be{d5-N} 
\frac{1}{M_L} \bY_\nu^{i j} = \bY^{T ik}_N{\bf M}^{-1 k l}_N\bY^{l j }_N    .
\ee
Finally, at the electroweak scale 
Majorana masses for the neutrinos are obtained,
\be{N-mass}
{\bf m}^{ij} _\nu = \frac{v_2^2}{M_L} \bY_\nu^{i j} =v_2^2~ 
\bY^{T ik}_N{\bf M}^{-1 k l}_N\bY^{l j }_N  .
\ee 

As a matter of fact, there is a certain degree of ambiguity in 
the effective neutrino mass matrix: its flavour structure 
reflects both that of  the arbitrary  Dirac-like 
matrix $\bY_N$  
and that of  the matrix ${\bf M}_N$.
From a bottom-up perspective, this implies that 
neutrino masses and mixing angles, inferred by 
the low-energy neutrino data \cite{analysis}, 
may only reflect the {\it effective} 
Yukawa matrix $\bY_\nu$ and the overall mass scale $M_L$ (modulo 
radiative corrections)  but cannot 
be unambiguously related to the more fundamental quantities, $\bY_N$ and 
${\bf M}_N$ (for a recent discussion on this aspect see for example 
 \cite{DI} and references therein). In other words, the low-energy 
parameters, described by $\bY_\nu$, 
which amount to 6 real parameters + 3 phases, are less than 
the number of the independent  `fundamental' physical parameters in $\bY_N$ 
and $\bM_N$, which instead are 12 real parameters + 6 phases.

Now, let us consider the triplet seesaw scenario. The relevant 
superpotential terms are
\be{L-T}
\frac{1}{\sqrt{2}}\bY^{ij}_{T} L_i T L_j  + 
\frac{1}{\sqrt{2}}\la_1 H_1 T H_1 +
\frac{1}{\sqrt{2}} \la_2 H_2 \bar{T} H_2 +  M_T T \bar{T} +
\mu H_1 H_2, ~~~~~~~
\ee
where the super-multiplets $T, \bar{T}$ are in a vector-like  
$SU(2)_W\times U(1)_Y$ 
representation\footnote
{Notice that in the supersymmetric 
picture two triplets, $T$ and $\bar{T}$, are required.
}, $T \sim (3,1)$ and $\bar{T} \sim (3,-1)$, 
and $H_1$ is the Higgs doublet 
with hypercharge $Y=-1/2$.
The matrix $\bY^{ij}_{T}$ is in general a  $3 \times 3$ 
symmetric matrix, 
$\bY^{ij}_{T} =\bY^{ji}_{T}$.  If we assign 
lepton number $L=-2~ (2)$ to the 
triplet $T ~(\bar{T})$, we can see that 
the $\la_1, \la_2$-couplings explicitly break 
$L$. If instead we assign $L= -2$  to $T$ and $L=0$ to $\bar{T}$, 
then the $L$-breaking parameters are $M_T$ and $\la_1$. 
In eq.~(\ref{L-T}) the triplets $T$, $\bar{T}$ are represented 
as  $2\times 2$ matrices, namely
\be{repT}
T = (i \sigma_2)\vec{T}\cdot {\vec{\sigma}}= 
\mat{{T^0}}{ -\frac{1}{\sqrt2}T^{+}} 
{-\frac{1}{\sqrt2}{T^+}}{-{T^{++}}} , ~~~~~~~~
\bar{T} = (i \sigma_2)\bar{\vec{T}}\cdot {\vec{\sigma}}= 
\mat{{\bar{T}^{- -}}}{ -\frac{1}{\sqrt2}{\bar{T}^{-}}} 
{-\frac{1}{\sqrt2}{\bar{T}^-}}{-{\bar{T}^{0}}} , 
\ee
where the matrices $\sigma_a ~ (a=1,2,3)$ are the Pauli matrices and 
the components $T_1, T_2, T_3$ and $\bar{T}_1, \bar{T}_2, \bar{T}_3$  
of $\vec{T}$ and $\bar{\vec{T}}$, respectively,  can be 
easily inferred\footnote{
As a consequence of the representation adopted for $T$ and $\bar{T}$ 
(\ref{repT}), the $SU(2)_W$-invariant mass term in eq.~(\ref{L-T}) 
is to be understood as $M_T T\bar{T} = M_T {\rm Tr}(T i\sigma_2 \bar{T}
i\sigma_2)$.}.
By decoupling the triplet states at the scale $M_T$ we again obtain 
the   $d=5$ effective operator of eq.~(\ref{d5}). 
Now, however, 
the flavour structure of the matrix $\bY_\nu$ is {\it the same} as 
that of $\bY_T$, that is:
\be{d5-T}
\frac{1}{M_L} \bY^{ij}_\nu = \frac{\la_2}{M_T} \bY^{ij}_T ,      
\ee
and the Majorana neutrino mass matrix is given by:
\be{T-mass}
{\bf m}^{ij}_\nu = 
\frac{v_2^2 }{M_L} \bY^{i j}_\nu=\frac{v_2^2 \la_2}{M_T} \bY^{i j}_T .
\ee 
Fig.~\ref{f1} shows the diagrams inducing the supersymmetric $d=5$ 
operator (\ref{d5}) in the $N$-seesaw (a) and $T$-seesaw (b).
In terms of component fields, the neutrino mass operator is generated 
through the exchange of the $N$ fermion component in the former realization, 
while 
in the latter realization it is generated 
by the exchange of the $T$ scalar component  and 
of the $\bar{T}$ F-component which gives rise  to the effective 
scalar coupling $\la_2 M_T  H_2 T^\dagger H_2$. 
This latter feature also allows  us to appreciate 
the supersymmetric version (\ref{L-T}) of the $T$-induced seesaw. 
In the non-supersymmetric case \cite{TS1} the coefficient 
of the cubic interaction $H T^\dagger H$ is an independent 
mass parameter, say $\Lambda$, 
and therefore $\bm_\nu = v^2 \frac{\Lambda}{M^2_T} \bY_T$.

\begin{figure}[htb]
\begin{center}
\begin{picture}(340,80)(0,0)
\Text(20,40)[]{(a)}
\Text(40,10)[bl]{$L$}
\Text(40,75)[tl]{$H_2$}
\Text(90,30)[]{$N$}
\Text(140,10)[br]{$L$}
\Text(140,75)[tr]{$H_2$}
\ArrowLine(50,20)(70,40)
\ArrowLine(50,60)(70,40)
\ArrowLine(90,40)(70,40)
\ArrowLine(90,40)(110,40)
\ArrowLine(130,20)(110,40)
\ArrowLine(130,60)(110,40)
\ArrowLine(270,0)(290,20)
\ArrowLine(310,0)(290,20)
\ArrowLine(290,40)(290,20)
\ArrowLine(270,80)(290,60)
\ArrowLine(310,80)(290,60)
\ArrowLine(290,40)(290,60)
\Text(240,40)[]{(b)}
\Text(260,0)[b]{$L$}
\Text(320,0)[b]{$L$}
\Text(260,80)[t]{$H_2$}
\Text(320,80)[t]{$H_2$}
\Text(300,30)[]{$T$}
\Text(300,50)[]{$\overline{T}$}
\end{picture}
\end{center}
\vspace{0.3 cm}
\caption{\small Contributions to the $L L H_2 H_2$ effective operator:
(a) heavy singlet exchange;  (b) heavy triplet exchange.}
\label{f1}
\end{figure}
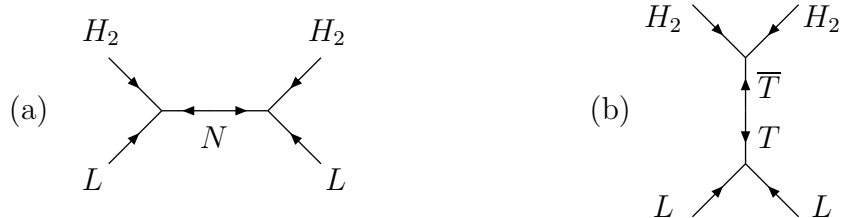
The difference with the standard $N$-induced seesaw case is manifest:
in the $T$-seesaw scenario the neutrino mass matrix  can be {\it directly} 
related to the fundamental $\bY_T$ matrix and to one relevant mass scale 
parameter, $M_T/\la_2$. 
In particular,  
the amount of lepton-flavour violation measured at low-energy 
through the lepton mixing can  directly be linked to the 
only  source of lepton-flavour violation, the  Yukawa matrix 
$\bY_T$ given at the scale $M_T$. 
The counting of the independent parameters reveals indeed that in 
this model the independent $\bY_T$ parameters, 6 real parameters + 3 phases, 
are just matched by the low-energy physical parameters. There are left other 
2 real parameters, $M_T$ and $\la_2$, whose ratio is  directly involved 
in the neutrino mass generation, plus $\la_1$ which in general 
is complex and is not directly connected to the neutrino parameters.   
In summary: while in the 
$N$-induced seesaw scenario there are two lepton-flavour violation 
sources, the matrices $\bY_N$ and $\bM_N$, 
in the $T$-induced seesaw scenario the matrix $\bY_T$ 
is  the only source of lepton-flavour violation\footnote{
Notice that in this seesaw scenario one triplet pair  $T, \ov{T}$ 
is enough to generate non-vanishing 
mass for all three neutrinos, whilst in the standard seesaw three singlets 
$N$ are necessary for that. 
Also notice that, from the flavour point of view, 
the seesaw realized by exchanging hyperchargeless 
triplets $T'\sim (3,0)$ \cite{ma} is more similar to the $N$-induced seesaw. 
Indeed: i) such triplets $T'$ are exchanged in the same channel as 
the singlets $N$ (see Fig.~\ref{f1} (a)); ii)  three hyperchargeless triplets 
are required  to give mass to all three neutrinos; iii) 
two sources of flavour-dependence appear, 
the matrix $Y_{T'}$ in the Yukawa interaction 
$ \bY_T L T' H_2$,    and the mass matrix $\bM_{T'}$ of the triplets.}.

The implications of all this are even more dramatic when we consider 
the lepton-flavour violation induced through renormalization effects.
Since a long it has been pointed out 
that the Yukawa couplings $\bY_N$ in eq.~(\ref{L-N}) 
can induce non-vanishing lepton-flavour violating  entries in the 
mass matrices of the left-handed sleptons through radiative 
corrections \cite{old}, 
even in the minimal SUSY scenario 
with universal soft-breaking terms at the GUT scale $M_G$, $m^2_{\tilde{L}}= 
m^2_{\tilde{e}}= \cdots = m^2_0\unity$. The form of the LFV entries is
\be{Nsusy}
 (\bm^{2 }_{\tilde{L}})_{ij} \propto
 m^2_0 (\bY^{\dagger}_N \bY_N)_{ij} {\log}\frac{M_G}{M_N} , ~~~~~~
i\neq j .
\ee 
In this scenario, according to our previous discussion, the size of LFV 
cannot be unambiguously predicted in a bottom-up approach 
making use of the low-energy data\footnote{The authors of \cite{DI} 
indeed regard the combination $\bY_N^\dagger \bY_N$, responsible 
for the LFV in the slepton masses, 
 as a further `observable' which provides the lacking 6 real parameters + 
3 phases necessary to fully determine both $\bY_N$ and $\bM_N$.}.

Now let us see what can occur in the $T$-seesaw scenario. In this case, 
the lepton-flavour violating entries are {\it directly }
connected to the effective neutrino mass matrices, as
\be{Tsusy1}
  ( \bm^{2 }_{\tilde{L}})_{ij} \propto  
 m^2_0 (\bY^{\dagger}_T \bY_T)_{ij} {\log}\frac{M_G}{M_T} , ~~~~~~
i\neq j,
\ee
or more explicitly,
\be{Tsusy2}
  (\bm^{2 }_{\tilde{L}})_{ij}\propto m^2_0 
\left(\frac{M_T}{\la_2 v^2_2}\right)^2 (\bm^{\dagger}_\nu \bm_\nu )_{ij}
{\log}\frac{M_G}{M_T}\sim 
m^2_0 \left(\frac{M_T}{\la_2 v^2_2}\right)^2
 \left[\bU (\bm^{D }_\nu)^2 \bU^\dagger\right]_{ij}
{\log}\frac{M_G}{M_T} .
\ee     
This expression enables us to {\it univocally} 
predict the ratio of the lepton-flavour 
violation in the 2-3 sector with that in the 1-2 sector, essentially 
in terms of the low-energy parameters,  namely:
\be{LF23-12}
 \frac{ ( m^{2 }_{\tilde{L}})_{\tau \mu}}
  {( m^{2 }_{\tilde{L}})_{\mu e} } \approx 
\frac{\left[\bU (\bm^{D }_\nu)^2\bU^\dagger\right]_{\tau \mu}}
{\left[\bU (\bm^{D }_\nu)^2\bU^\dagger\right]_{\mu e}} , 
\ee
or for the 1-3 sector
\be{LF13-12}
 \frac{ ( m^{2 }_{\tilde{L}})_{\tau e}}
  {( m^{2 }_{\tilde{L}})_{\mu e} } \approx 
\frac{\left[\bU (\bm^{D }_\nu)^2\bU^\dagger\right]_{\tau e}}
{\left[\bU (\bm^{D }_\nu)^2\bU^\dagger\right]_{\mu e}} .
\ee
Thus we can relate the rate of the  $\tau \to \mu \gamma$ or 
  $\tau \to e \gamma$ decay with 
that of the  $\mu \to e \gamma$ decay. 
This is the main feature of our discussion.  

This scenario is susceptible of being further elaborated. Indeed, the 
presence of the extra $SU(2)_W$-triplet states $T, \bar{T}$ 
at intermediate energy would spoil the gauge coupling unification 
which takes place with the field content of the MSSM. 
A simple way to save 
 gauge coupling unification is  to introduce 
  more states $X$, 
to complete a certain representation $R$ -- such that $R = T +X$ -- 
 of some unifying gauge group $G$, 
$G \supset SU(3)\times SU(2)_W\times U(1)_Y$. 
Thus   we can envisage three (minimal) scenarios:
\begin{description}
\item
(A)- The states $X$, though with mass  $M_X \sim M_T$, are assumed 
to have vanishing or negligible  
interactions with the other states, except for the gauge interactions. 
This may be the case either if we prefer not to embed the theory 
in any definite grand unified theory (GUT) 
or if we embed it in some  GUT and the Yukawa interactions 
of the states $T$ and $X$ 
have different strength due to GUT breaking effects. 
In particular, we could have negligible Yukawa coupling 
for the fragments $X$ and, on the contrary, 
non-vanishing Yukawa coupling for the fragments $T$.    
\item
(B)- The theory is embedded in a GUT but, contrary 
to the ansatz (A), the Yukawa couplings of the states $X$ 
are assumed to be non-vanishing and related to those of the triplet partners 
$T$. Indeed, this is generally  the case in  minimal 
GUT models.  
In this case we will generate not only lepton- flavour violation 
but also closely related  
flavour violation  in the quark sector (related to the 
$X$-couplings).
\item
(C)- There are no extra states $X$ or, equivalently, they are considered 
to be split in mass from the triplets $T$ and be decoupled 
at a scale $\mu \geq M_G$. In this case, the simple  unification of 
gauge couplings  
is lost and large threshold corrections are needed to recover it. 
\end{description}


\section{SUSY $SU(5)$  scenario with $SU(2)_W$ triplets}
As already mentioned in the previous section, the extra states $T, \bar{T}$ 
with mass $M_T$ much below the GUT scale would destroy the gauge coupling 
unification. 
In principle the latter property could  be recovered at the price of 
invoking large threshold corrections.
In the next, however, we prefer to maintain the simple 
gauge coupling unification.
To this purpose, the  field content of the model can be minimally 
extended by adding the  other components of the $SU(5)$ representations, 
15 and $\ov{15}$,  in which the triplets $T$ and $\bar{T}$  
can indeed fit.
In terms of $SU(3)\times SU(2)_W\times U(1)_Y$ representations, 
the 15-multiplet decomposes as follows:
\beqn{quin}
&& 15 =  S + T + Z , \nonumber \\
&& S\sim (6,1,-\frac23) , ~~~~  T\sim (1,3,1) , ~~~~Z \sim (3,2,\frac16)  ,
\eeqn
(the $\ov{15}$-decomposition is obvious). 
The presence of these extra states fitting a complete GUT 
multiplet  changes  the value of the gauge coupling $\alpha_G$ at 
the GUT scale, with respect to the MSSM case, 
but does not modify the value of the unification scale $M_G$ (to one 
loop accuracy). 
The $\beta$-functions of the gauge couplings in the RGEs get modified 
as follows $(a=1,2,3)$:
\beqn{betas}
&&16\pi^2 \frac{d g_a}{dt} =  B_a  g^3_a , \nonumber \\
&& B_1 = b_1 + \frac35(\frac83 n_S + 3 n_T   +\frac{1}{6} n_Z) = 
b_1 +\frac72 n_{15} , \nonumber \\
&& B_2 = b_2 + 2 n_T + \frac32 n_Z= b_2 +\frac72 n_{15} , \nonumber \\
&& B_3 = b_3 + \frac52 n_S + n_Z = b_3 +\frac72 n_{15} , 
\eeqn
where $b_a$ are the coefficients of 
the $\beta$-functions in the MSSM, namely $b_1=\frac{33}{5}, b_2=1, b_3=-3$,  
and we have  explicitly shown the contribution of the new states 
($n_S =N_S + N_{\bar{S}}$, $n_T=N_T + N_{\bar{T}}$, 
$n_Z=N_Z + N_{\bar{Z}}$ and $n_{15}=N_{15} + N_{\bar{15}}$ 
in a self-explanatory notation).
As expected, for each $B_a$  
the overall contribution of $S,T,Z$ just reproduces  the Dynkin 
index $\frac72$ of the $SU(5)$ 15 representation.
The enhancement of the $\beta$-functions makes the gauge couplings 
increase faster at higher energy. For instance, 
by using the low-energy 
values of $\alpha_{\rm em}$, 
$\alpha_{\rm s}$ and $\sin^2\theta_W$ and 
assuming 
an average SUSY threshold close to the top mass, for   $T,S,Z$ masses  
around $10^{14}~{\rm GeV}$ we find that  at one-loop 
$g_1$ and $g_2$ get unified at $M_G\sim 2\times 10^{16}~{\rm GeV}$ 
to the common value $g_G\sim 0.88$, while the value of $g_3$ differs 
by one per mill or so.
In the MSSM we would find $g_G\sim 0.71$. 

The $SU(5)$ invariant superpotential 
(omitting for simplicity  the flavour indices) reads as:
\beqn{su5}
W_{SU(5)}&\!\!\!=\!\!\!& \frac{1}{\sqrt2} \bY_{15} \bar5~  15 ~\bar5    + 
 \frac{1}{\sqrt2} \la_1  ~\bar{5}_H~ 15~ \bar{5}_H 
+  \frac{1}{\sqrt2}~ \la_2  {5}_H ~\ov{15}~ {5}_H + 
 \bY_5 10  ~\bar5 ~ \bar5_H  \nonumber \\  
&& + \bY_{10} 10 ~10  ~ 5_H + M_{15} 15~ \ov{15}+ M_5 \bar5_H~ 5_H , 
\eeqn
where  the matter multiplets are understood as 
$\bar5  = (d^c, L)$, $10 = (u^c,e^c,Q)$ and the Higgs doublets 
fit with their coloured partners, $t, \bar{t}$ as 
${5}_H = (t, H_2), \bar{5}_H = (\bar{t}, H_1)$. 
In the 15-multiplet  the states $S, T$ and $Z$ are accommodated as:
\be{su5d}
15^{A B} = \mat{{S}^{ a b}}{ \frac{1}{\sqrt2}{Z^{a j}}}
{\frac{1}{\sqrt2}{Z^{ bi}}}{ {T}^{ij}}
\ee
where the $SU(5)$-indices $A, B =1,2,3,4,5$ are decomposed into 
$SU(3)$ indices $a,b =1,2,3$ and $SU(2)_W$ indices $i,j= 4,5$, $A= (a,i), 
B=(b,j)$. In eq.~(\ref{su5d}) it is understood that 
${S}^{aa} = \hat{S}^{a a}$, 
${S}^{ab} =  
\frac{1}{\sqrt2} \hat{S}^{a b}$ ($a\neq b$) and 
${T}^{ii} =\hat{T}^{ii}$, 
${T}^{ij} =  
\frac{1}{\sqrt2}\hat{T}^{i j}$  ($i\neq j$) (cfr. eq.~(\ref{repT})) 
where the fields $\hat{S}^{ab}, \hat{T}^{i j}$ are those canonically 
normalized.
It is well-known that the minimal $SU(5)$ model in which the 
Yukawa matrices $\bY_5, \bY_{10}$ 
 are true constants is not  phenomenologically 
satisfactory. The latter should be rather understood as 
field dependent quantities, e.g. 
$\bY_5 (\Phi) = \bY^{(0)}_5 + \bY_5^{(1)} {\Phi}/{M} + \cdots$ where 
$\Phi$ is the adjoint 24 of $SU(5)$ and 
$M$ is some cutoff scale larger than the GUT scale   
$M_G$. This perspective allows us to 
correct
certain $SU(5)$-symmetry relations, 
such as $\bY_d=\bY^T_e$ \cite{su5}.  
Moreover, some mechanism is also necessary to split the 
masses\footnote{
Also for the minimal `technical' realization of the doublet-triplet 
splitting \cite{DT} we can give an 
interpretation of the mass parameter $M_5$,  analogous to that 
adopted for the Yukawa couplings, i.e. 
$M_5(\Phi) = M^0_5 +\la_5 \Phi +\cdots$.}  
of the doublet $H_{1,2}$ and triplet $t, \bar{t}$ 
components of $5_H, \bar{5}_H$ not to 
have fast proton decay mediated by the coloured states 
$t, \bar{t}$ \cite{DT}.  
We shall therefore adopt this 
point of view for  the whole $SU(5)$ 
extended model of eq.~(\ref{su5}).
In the   $SU(5)$-broken phase, beneath $M_G$, the superpotential reads as:  
\beqn{su5b}
&&\frac{1}{\sqrt2} ( \bY_T L T L +   \bY_S  d^c S d^c ) 
+ \bY_Z  d^c  Z L + \bY_d d^c Q H_1   +\bY_e e^c L H_1 
+ \bY_u u^c Q H_2 \nonumber \\
&&     +
\frac{1}{\sqrt2}(\la_1 H_1 T H_1 +\la_2 H_2 \bar{T} H_2)  
 + M_T T\bar{T}+ M_Z Z\bar{Z}+M_S S\bar{S} + \mu H_1 H_2  .
\eeqn 
As  already mentioned, 
the couplings involving the coloured triplets $t, \bar{t}$ 
do not appear as the latter are assumed to decouple at the GUT scale $M_G$ 
to suppress dangerous $B-L$ violating $d=5$ operators.  
On the contrary, at the decoupling 
of the 15 fragments, no $B-L$ violating  $d=5$ operators are induced, 
apart from the `neutrino' operator, 
since the  $\ov{15}$ states do not couple to the matter 
multiplets $\bar{5}, 10$. 
Only flavour-conserving $d=6$ operators are generated, i.e.  
$(L d^c)(\bar{L} \bar{d}^c),~ (d^c d^c)(\bar{d}^c \bar{d}^c), ~  
(L L)(\bar{L} \bar{L})$ in the K\"ahler potential,  
which, being suppressed by the square of the large  scale 
$M_{15}$, are not relevant for the low-energy phenomenology.
Notice that in the minimal case the masses $M_T, M_S, M_Z$ are equal 
at the GUT scale,
\be{unif}
 M_S = M_T =M_Z = M^0_{15} .
\ee
However, this unification 
relation could be modified e.g. due to $\Phi$-insertions, 
as mentioned above\footnote{Moreover, even if eq.~(\ref{unif}) 
holds at $M_G$,  renormalization effects split the masses at lower 
energies. However,  the relative splitting is not large and 
we will decouple all the components $T,S,Z$  at the common 
threshold scale $M_T$.}.
Similarly, the unification of the Yukawa couplings 
at the GUT scale,
 \be{unif1}
 \bY_S  = \bY_T =\bY_Z = \bY^0_{15} , 
\ee
could either hold or not.
Finally, we stress  that in this $SU(5)$ framework  the flavour violation 
is encoded not only in  $\bY_T$ (as already elucidated above) 
but also in  $\bY_S$ and $\bY_Z$. 
Therefore, non-vanishing barion-flavour violating entries 
in the mass matrix $\bm^2_\td$ 
of the sdown squarks $\td$ are induced by radiative corrections. 
In summary, the three scenarios (A), (B) and (C),  
put forward in Section 2, can be rephrased as follows:
\begin{description}
\item
(A)- All fragments 
$T, S$ and $Z$ have the same mass, i.e. eq.~(\ref{unif}) holds at $M_G$.
However, the couplings $\bY_S, \bY_Z$ are assumed to be negligible, i.e. 
eq.~(\ref{unif1}) does not hold.
In this case 
only the interactions with the triplets $T$ and so the couplings $\bY_T$ 
drive the lepton-flavour violation in the slepton scalar masses.

\item
(B)- Both the masses and the Yukawa couplings of the 15 states are unified, 
i.e. both eqs.~(\ref{unif}) and (\ref{unif1}) hold.
Therefore 
all the couplings $\bY_S, \bY_T, \bY_Z$ will induce flavour violation 
in both the slepton $\tl$ and squark $\td$ masses. 

\item
(C)- The triplet mass $M_T$ is much smaller than $M_S$ and $M_Z$, which are 
${\cal O}(M_G)$. This could be achieved for instance by tuning 
the coefficients of the singlet and adjoint components of the $15~\ov{15}$ 
`mass', in analogy to what is done for the  
doublet-triplet splitting  for   $5_H, \bar5_H$.
\end{description}

In the following we shall focus on (A) and (B) and disregard the case (C).

\section{$\bY_T$ from neutrino masses and mixing angles} 
Now, we relate the low-energy parameters with the relevant neutrino 
Yukawa couplings by adopting a bottom-up criterion. 
By decoupling the states $T,\bar{T}$, 
the $d=5$ effective operator emerges 
\be{d5-T2}
 \frac{\la_2}{2 M_T} \bY^{ij}_T (L_i H_2)  (L_j H_2) , 
\ee 
where the matrix $\bY_T$, through the matching of eq.~(\ref{d5-T}), 
can be connected to $\bY_\nu$ which parameterizes the usual 
$d=5$ operator (\ref{d5}) 
\footnote{
The expression given in eq.~(\ref{d5-T2}) is the leading contribution 
to the neutrino mass operator. 
This  arises from the $\bar{T}$ F-term  scalar interaction, 
$\lambda_2 M_T T^\star H_2 H_2$. The  $H_1$ F-term  scalar interaction, 
$\mu \lambda_1 H_2^\star T H_1$, gives rise to the sub-leading contribution 
$\lambda_1 (\mu/M_T) (v_2 v_1 )/M_T$.}. 
 
We recall that the data 
from solar and atmospheric neutrino 
experiments concern the neutrino mass eigenvalues $m_1,m_2, m_3$ 
and mixing angles. Therefore, in the basis 
in which the Yukawa 
matrix $\bY_e$ 
is diagonal, eqs.~(\ref{nu-mass}) and (\ref{mix}) allow us 
to determine the   coupling matrix $\bY_\nu/M_L$  at low-energy 
and then, taking into account the running up to $M_T$ \cite{RG-nu}, 
 also at the scale 
$M_T$.\footnote{The radiative corrections from the electroweak scale up to 
 $M_T$ are not important in the case of hierarchical neutrino mass 
spectrum and  amount to an overall common factor 
\cite{RG-rec}. 
Nevertheless these effects are incorporated in the numerical 
analysis.} 
The mixing matrix $\bU$ is parameterised in the standard way:
\be{U}
\bU= \matr{c_{12}c_{13}}{s_{12}c_{13}}{s_{13}e^{-i\delta}}
{-s_{12}c_{23}-c_{12}s_{23}s_{13}e^{i\delta}}
{c_{12}c_{23}-s_{12}s_{23}s_{13}e^{i\delta}} {s_{23}c_{13}} 
{s_{12}s_{23}-c_{12}c_{23}s_{13}e^{i\delta}}
{-c_{12} s_{23}-s_{12}c_{23}s_{13}e^{i\delta}} {c_{23}c_{13}}
\ee
where $s_{ij}$ and $c_{ij}$ are the cosine and sine respectively of the 
three mixing angles $\theta_{12}, \theta_{23}, \theta_{13}$ and $\delta$ is 
the CP-violating phase which in the following is neglected for simplicity.
As for the phenomenological input, we assume maximal 2-3 mixing, 
$\theta_{23}= 45^\circ$, as required by the atmospheric neutrino data 
\cite{analysis}.
 We consider  hierarchical neutrino mass 
spectrum, $m_1\ll m_2\ll m_3$, such that it reads as:
\be{numass}
m_1 \approx 0, ~~~~~~m_2 = (\Delta m^2_{\rm sol})^{1/2} , ~~~~~~
m_3 = (\Delta m^2_{\rm atm})^{1/2} ,
\ee
where we take $\Delta m^2_{\rm atm} \sim 3\times 10^{-3} {\rm eV}^2$. 
As regards the solar neutrino case, the most favoured range 
for $\Delta m^2_{\rm sol}$ is that selected by the large-mixing angle 
(LMA) solution, 
 $\Delta m^2_{\rm sol} \sim 6\times 10^{-5} {\rm eV}^2$ \cite{analysis}. 
The corresponding best fit value for the mixing angle 
is $\theta_{12}\sim 33^\circ$. 
However, for the 
sake of discussion in the following we bear in mind also different 
possibilities, such as the typical values  
$\Delta m^2_{\rm sol} \sim 5\times 10^{-6} {\rm eV}^2$ and 
$\theta_{12}\sim 4^\circ$  
of the small-mixing angle (SMA) solution. 
Taking also into account the 
CHOOZ limit, $\sin\theta_{13} < 0.1$  \cite{chooz}, 
we set $\theta_{13}=0$ for 
simplicity and later comment also on the non-zero $\theta_{13}$ case.
Then at low-energy the symmetric 
matrix $\bY_\nu$ appears as:
\be{ynu-low}
{\bY_\nu} = \frac{M_L}{v^2_2} \matr
{m_2 s^2_{12}}{\frac{1}{\sqrt{2}}{m_2} c_{12} s_{12} }
{-\frac{1}{\sqrt{2}}{m_2} c_{12} s_{12} }
{} {\frac{1}{{2}}({m_3} + {m_2} c^2_{12}) }
 {\frac{1}{{2}}({m_3}- {m_2} c^2_{12} )}
{}{} {\frac{1}{{2}}({m_3}+ {m_2} c^2_{12}) } .
\ee 
By considering the phenomenological inputs, we observe that the entries 
of the 2-3 sector are all comparable and can be of order one (essentially 
irrespectively 
of the $\theta_{12}$ value)  if the  
 scale $M_L$ is close to $M_G \sim 10^{16} {\rm GeV}$. 
The remaining entries are one order of magnitude smaller than 
those of the 2-3 sector, 
for  $\theta_{12}$ in the LMA range, 
while they are much smaller for $\theta_{12}$ 
in the SMA range. Up to an overall  factor due to 
the renormalization effect, the structure 
obtained for $\bY_\nu$ is finally transferred to
 $\bY_T$ at the scale $M_T$ according to eq.~(\ref{d5-T}). 
 We stress again that in this  $T$-seesaw scenario the 
bottom-up approach here adopted is the most general one since  
 the structure  of $\bY_T$ is  unambiguously fixed  by the 
experimental data 
themselves. What can be left to our choice is the overall scale 
$M_T/\la_2$.
We shall take $M_T$ in the range $10^{11}\div 10^{15} ~{\rm GeV}$ 
and vary $\la_2$ in an appropriate range.

\section{Lepton-flavour violation in the soft-breaking terms}
The general soft SUSY-breaking terms in our model are given as\footnote{
For the sake of simplicity, in the following we disregard 
all what concerns the up-quark sector parameters, such as the scalar masses 
$\bm^2_{\tilde{Q}}$ etc. because they do not enter directly our discussion.}:
\beqn{Lsoft} 
-{\cal L}_{\rm soft}&\!\!\! \!\!=\!\!\!\!\! &   \tl^\dagger \bm^2_{\tl} \tl +
 \te \bm^2_{\te }  \te^\dagger+  \td  \bm^2_{\td} \td^\dagger + 
 m^2_{H_1} H^\dagger_1 H_1+ m^2_{H_2} H^\dagger_2 H_2  \nonumber \\
&& +
( H_1 \te \bA_{e } \tl  + H_1 \td \bA_{d } \tq+
\frac12 M_a \tilde{\la}_a \tilde{\la}_a +  
B\mu H_1 H_2 + {\mbox h.c.})   \nonumber \\ 
&&  +
\left[\frac{1}{\sqrt2}( \bA_{T}  \tl T \tl +\bA_{S}  \td S \td) +
\bA_{Z}  \td Z  \tl  + \frac{1}{\sqrt2}(A_1  H_1 T H_1  + A_2   H_2 
\bar{T}  H_2) 
\right. \nonumber \\
&& \left.+ B_T M_T T \bar{T} +  
B_S M_S S \bar{S}+ B_Z M_Z Z \bar{Z}+
 {\rm h.c.}\right]  \nonumber \\  
&& + m^2_T T^\dagger T+ m^2_{\bar{T}} \bar{T}^\dagger \bar{T}  +
 m^2_S S^\dagger S+ m^2_{\bar{S}} \bar{S}^\dagger \bar{S}  +
m^2_Z Z^\dagger Z+ m^2_{\bar{Z}} \bar{Z}^\dagger \bar{Z}  
\eeqn
where we have shown 
in the first lines the relevant MSSM terms, 
according to the standard notation (the 
 soft mass terms for the sleptons, squarks, Higgs bosons, the trilinear terms 
and the gaugino masses $M_a$),  
while in last lines  
we have collected the new terms involving $T,S,Z$.
In the following we assume at the GUT scale, irrespectively of 
the scenario (A) or (B)\footnote{
These universal boundary conditions are mainly motivated 
by simplicity and are not assumed for other soft parameters.
For instance, we do not declare 
the soft-breaking mass of the doublet scalar $H_{2}$,  
as it does not directly enter in our analysis.
In this way the $\mu$-parameter in the 
superpotential is not constrained by the electroweak 
radiative-breaking condition and will be fixed independently. 
We would like to make clear, however, that 
the aim of this work is to present the global features of the 
SUSY $T$-induced seesaw. Therefore,  
the choice of these initial 
conditions, as well as of other parameters such as $\tan\beta$ and 
 $\mu$, which enter in the computations of the decay rates, is 
only made for illustrative purposes.
}:
\beqn{univ}
&& \bm^2_{\tl} = \bm^2_{\te}=\bm^2_{\td} = m^2_0 \unity , ~~~~~ 
\nonumber \\
&&
m^2_S=m^2_{\bar{S}}=m^2_T=m^2_{\bar{T}}=m^2_Z=m^2_{\bar{Z}} 
=m^2_{H_1}=m^2_0 ,  \nonumber\\
&&
\bA_e = A_0 \bY_e , ~~~~~~\bA_d= A_0\bY_d,~~~~~
A_1 = A_0 \la_1, ~~~~~~ A_2 = A_0 \la_2, \nonumber \\ 
&& M_1=M_2=M_3= M_g, 
\eeqn
where $m_0$ is the universal scalar mass, $A_0$ is the universal 
mass parameter for the trilinear terms  and $M_g$ is the common gaugino mass. 
As for the remaining trilinear couplings, their GUT conditions 
are in the scenario (A):
\be{tri-A}
\bA_T = A_0 \bY_T, ~~~~~~\bA_S=\bA_Z =0 , 
\ee
and in the scenario (B)
\be{tri-B}
\bA_T = \bA_S=\bA_Z =A_0 \bY_T . 
\ee
The matrix $\bY_T$ which appears in eqs.~(\ref{tri-A}-\ref{tri-B}) is 
understood to be  evaluated at the GUT scale.
Once $\bY_T$ is determined at the scale $M_T$ by the low-energy data 
as we have seen above, its evolution from $M_T$ up to  $M_G$ 
is given by: 
\beqn{rgeYt}
16\pi^2 \frac{d \bY_T}{dt} & =& \bY_T\left( -\frac95 g^2_1 - 7 g^2_2 
+ \bY^\dagger_e \bY_e + 6\bY^\dagger_T\bY_T +{\rm Tr}
(\bY_T\bY^\dagger_T) + 3 \bY^\dagger_Z \bY_Z +|\la_1|^2\right)   \nonumber \\
&&+ (\bY^T_e \bY^\star_e  +3 \bY^T_Z \bY^\star_Z )\bY_T .
\eeqn
(See also Appendix.) 
In the scenario (B) the $\bY_T$ RGE is also coupled to those of 
$\bY_S, \bY_Z$ for which the initial conditions at the scale $M_T$ 
are determined iteratively under the constraints of eq.~(\ref{unif1}).
Below $M_G$ the universal pattern (\ref{univ}) of 
$\bm^2_{\tl}, \bm^2_{\td}$ etc. 
is spoiled by radiative effects induced by $\bY_T, \bY_S, \bY_Z$.
Then we have to evaluate the soft-breaking parameters at low-energy 
by solving the corresponding RGEs. These have been computed at one-loop 
and  collected in  Appendix.
In the leptonic sector we need to know the SUSY breaking matrices  
$\bm^2_\tl, \bm^2_\te$ and $\bA_e$ to finally compute the LFV decay rates.
They all receive flavour blind corrections from the gauge interactions 
which do not alter the flavour-conserving  structure 
they have at the GUT scale (see eq.~(\ref{univ})). 
However, they can acquire LFV entries (i.e. off-diagonal entries) 
if they get radiative corrections from the LFV Yukawa-matrices  $\bY_T, 
\bY_Z$. 
In the leading-log approximation, and neglecting radiative 
corrections induced by $\bY_e, \bY_d$, in the picture (A)  
the LFV entries at low-energy are given by ($i\neq j$):
\beqn{log}
(\bm^2_{\tl})_{ ij}& \approx& \frac{-1}{8\pi^2} (9 m^2_0 + 3 A^2_0) 
(\bY^\dagger_T\bY_T)_{ij} \log\frac{M_G}{M_T} ,  \nonumber\\
(\bm^2_{\te})_{ ij} &\approx& 0 ,  \nonumber \\
(\bA_{e})_{ ij}& \approx& \frac{-9}{16\pi^2}  A_0 
(\bY_e \bY^\dagger_T\bY_T)_{ij} \log\frac{M_G}{M_T} . 
\eeqn
and in the picture (B) we find:
\beqn{logB}
(\bm^2_{\tl})_{ ij}& \approx& \frac{-1}{8\pi^2} (18 m^2_0 + 6 A^2_0) 
(\bY^\dagger_T\bY_T)_{ij} \log\frac{M_G}{M_T} , \nonumber \\
(\bm^2_{\te})_{ ij} &\approx& 0 ,  \nonumber \\ 
(\bA_{e})_{ ij}& \approx& \frac{-9}{8\pi^2}  A_0 
(\bY_e \bY^\dagger_T\bY_T)_{ij} \log\frac{M_G}{M_T} , 
\eeqn
and in the squark sector:
\beqn{logBq}
(\bm^2_{\td})_{ ij}& \approx& \frac{-1}{8\pi^2} (18 m^2_0 + 6 A^2_0) 
(\bY^\dagger_T\bY_T)_{ij} \log\frac{M_G}{M_T} , \nonumber\\ 
(\bA_{d})_{ ij}& \approx& \frac{-9}{8\pi^2}  A_0 
( \bY^\dagger_T\bY_T \bY_d)_{ij} \log\frac{M_G}{M_T} , 
\eeqn 
where  we have taken into account the 
$SU(5)$-universality for the Yukawa matrices (\ref{unif1}).
So the mass matrix $\bm^2_\te$ remains diagonal and then flavour-conserving, 
while both  $\bm^2_\tl$ and $\bA_e$ acquire LFV elements. 
 Once the low-energy neutrino observables 
are fixed, the magnitude of 
these LFV elements will depend on the matrix $\bY^\dagger_T \bY_T \sim 
(M_T/\la_2 v^2_2)^2 \bm^\dagger_\nu \bm_\nu $, that is on the 
triplet mass threshold $M_T$ and on the coupling constant $\la_2$. 
Thus the relative size  of LFV 
in the 2-3 sector and 1-2 sector can be approximately predicted in terms 
of only the low-energy observables, as we anticipated in eq.~(\ref{LF23-12}). 
Such a ratio can be rewritten more explicitly: 
\be{predi}
 \frac{ (\bm^{2 }_{\tilde{L}})_{\tau \mu}}
  {(\bm^{2 }_{\tilde{L}})_{\mu e} } \approx \left(
\frac{m_3}{m_2}\right)^2 \frac{\sin 2\theta_{23}} {\sin 2\theta_{12}
\cos\theta_{23}} \sim 80 ,
\ee
where for the estimate we have taken  for $\theta_{12}$ and 
$m_2$ the values  selected by the LMA solution. 
For the case of SMA solution that ratio increases to $10^3 -10^4$. 
These results clearly hold in both  scenarios (A) and (B). 
Therefore, this estimate can 
 directly be translated into a prediction for the ratio of the decay rate 
of $\tau \to \mu \ga$ and $\mu\to e \ga$ as we shall show in the next Section.
We also recall  that since $\theta_{23} \sim 45^\circ$, 
the first generation is mixed with a state which is an equal 
(and indistinguishable) mixture of 
 the flavour states $\nu_\mu, \nu_\tau$. It is not surprising therefore that
the  ratio ${ (\bm^{2 }_{\tilde{L}})_{\tau e}}/
  {(\bm^{2 }_{\tilde{L}})_{\mu e} }$ be of order one:
\be{predi31}
 \frac{ (\bm^{2 }_{\tilde{L}})_{\tau e}}
  {(\bm^{2 }_{\tilde{L}})_{\mu e} }
 \approx 
\tan\theta_{23} \sim 1 . 
\ee
Notice also that the size of the lepton-flavour 
violating entries $(\bm^2_\tl)_{ij}$ 
 is about a factor 2 larger in the scenario (B) due to the 
extra contribution driven by the Yukawa couplings $\bY_{Z,S}$. 
This implies that the related decay-rates are further enhanced by a factor 
4 in the scenario (B).
Moreover, in the scenario (B), we have similar predictions 
for the sdown sector, namely:
\be{predi-dc}
 \frac{ (\bm^{2 }_{\td })_{b s}}
  {(\bm^{2 }_{\td })_{s d} } \approx \left(
\frac{m_3}{m_2}\right)^2 \frac{\sin 2\theta_{23}} {\sin 2\theta_{12}
\cos\theta_{23}} \sim 80 , ~~~~~~
\frac{ (\bm^{2 }_{\td })_{b d}}
  {(\bm^{2 }_{\td })_{s d} } \approx 
\tan\theta_{23} \sim 1 ,  
\ee
which show that the relative flavour-violation in the lepton and 
quark sector should be comparable in magnitude.

Finally, we would like to comment also upon the case in which the 1-3 
mixing is restored in the lepton mixing matrix. So for discussion
 we consider the present upper bound on 
$\theta_{13}$ \cite{chooz} and take $\sin\theta_{13}=0.1$. 
This leads to an enhancement of the LFV 
entries $(\bm^2_\tl)_{ \mu e} $ 
and $(\bm^2_\tl)_{\tau e} $ driven by 
the largest mass $m_3$, with respect to the $\theta_{13}=0$, namely
\beqn{mue13}
&&  \frac{ (\bm^{2 }_{\tilde{L}})_{ \mu e}|^{\theta_{13}\neq0}}
  {(\bm^{2 }_{\tilde{L}})_{\mu e}|^{\theta_{13}=0} } \approx 
1 + \left(
\frac{m_3}{m_2}\right)^2 \frac{\sin \theta_{13}} {\sin \theta_{12}
\cos\theta_{12}}  \sim 10 , \nonumber \\
&&  \frac{ (\bm^{2 }_{\tilde{L}})_{ \tau e}|^{\theta_{13}\neq0}}
  {(\bm^{2 }_{\tilde{L}})_{\tau e}|^{\theta_{13}=0} } \approx 1 -
\left(
\frac{m_3}{m_2}\right)^2 \frac{\sin \theta_{13}} {\sin \theta_{12}
\cos\theta_{12}} \sim - 10 .
\eeqn
Therefore, since the entry $(\bm^{2 }_{\tilde{L}})_{ \tau \mu}$ 
is not modified, the ratio of eq.~(\ref{predi}) becomes $\sim 7$, while 
that in eq.~(\ref{predi31}) remains the same. 
By considering the values for $m_2$ and $\theta_{12}$ 
of  the SMA solution, we find for $\sin\theta_{13}=0.1$  
a much  stronger relative enhancement, 
\beqn{mue13-sm}
&&  \frac{ (\bm^{2 }_{\tilde{L}})_{ \mu e}|^{\theta_{13}\neq0}}
  {(\bm^{2 }_{\tilde{L}})_{\mu e}|^{\theta_{13}=0} } \approx  
10^3 , \nonumber\\
&&  \frac{ (\bm^{2 }_{\tilde{L}})_{ \tau e}|^{\theta_{13}\neq0}}
  {(\bm^{2 }_{\tilde{L}})_{\tau e}|^{\theta_{13}=0} } \approx 
- 10^3 .
\eeqn
Analogous results 
are obtained for the entries  $(\bm^{2 }_{\td})_{ij}$. 

\section{The $\ell_i\to \ell_j \ga$ decay rates}
Let us briefly recall  here some points  related to the computation of 
the $\ell_i \to \ell_j \ga$ decay rate. 
The effective operator responsible for such a decay can be parameterised as 
\be{dipol}
{\cal L}_{\rm eff}= i g_e  m_i 
(C^{ij}_L \bar{l_j}\bar{\sigma}^{\mu \nu} \bar{l_i}^c + 
 C^{ij}_R {l_j}^c {\sigma}^{\mu \nu} {l_i})F_{\mu \nu} ,
\ee
where $g_e$ is the electromagnetic coupling and we use two-component 
spinor notation.
This leads to the branching ratio
\be{BR}
BR( \ell_i \to \ell_j \ga) = \frac{48 \pi^3 \alpha_{\rm em}}{G^2_F}\left(
|C^{ij}_L|^2 + |C^{ij}_R|^2 \right)BR(\ell_i \to \ell_j \nu_i \bar{\nu}_j) , 
\ee
where in the specific cases 
the lepton-flavour conserving 
branching ratio are 
$BR(\mu \to e \nu_\mu \bar{\nu}_e)\approx 1$, 
$BR(\tau \to \mu \nu_\tau\bar{\nu}_\mu)\approx 17\%$ and 
$BR(\tau \to e \nu_\tau\bar{\nu}_e)\approx 18\%$.
For the numerical analysis 
we have taken into account all contributions 
involving one-loop slepton-chargino and slepton-neutralino exchange, 
by using the complete  
formulas given for example in ref.~\cite{new}. 
Analogously to   the case
of the MSSM with singlets $N$ \cite{new}, 
 also in this scenario the 
main contributions come from $\tan\beta$-enhanced diagrams 
with chargino exchange.
In the mass-insertion approximation,  
we recall that the parameter dependence of $C^{ij}_L$ (the dominant 
coefficient) is 
\be{BRmain}
C^{ij}_L 
\sim 
\frac{g^2}{16 \pi^2}
\dfrac{(\bm^2_{\tl})_{ ij}} {\tilde{m}^4 } \tan\beta,  
\ee
where $\tilde{m}$ is  an average soft mass.
\begin{figure}[p]
\vskip -2.cm
\hglue -0.8cm
\epsfig{file=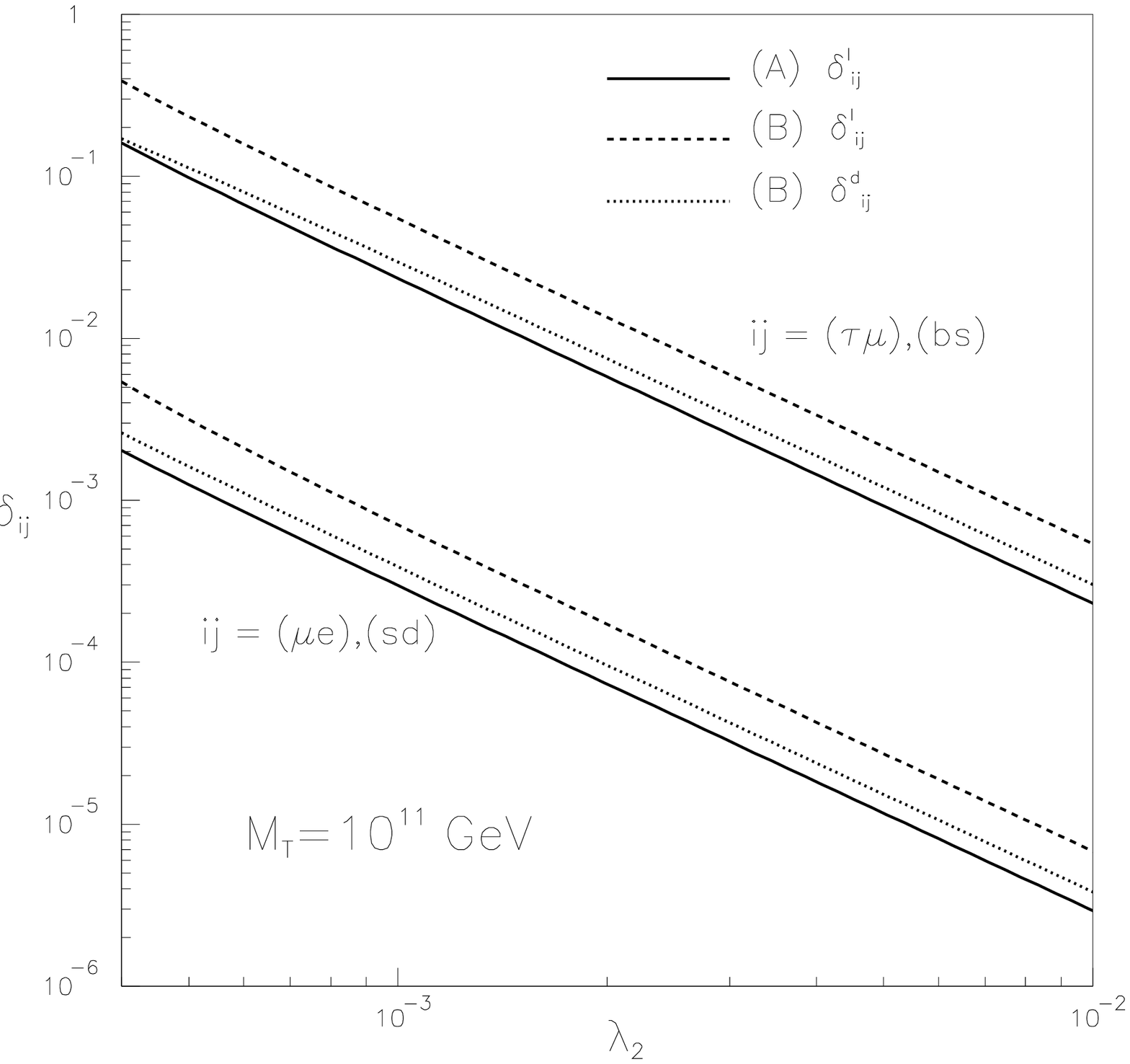,height=9.cm,width= 8.6cm}
\vglue -9.cm
\hglue 7.6cm
\epsfig{file=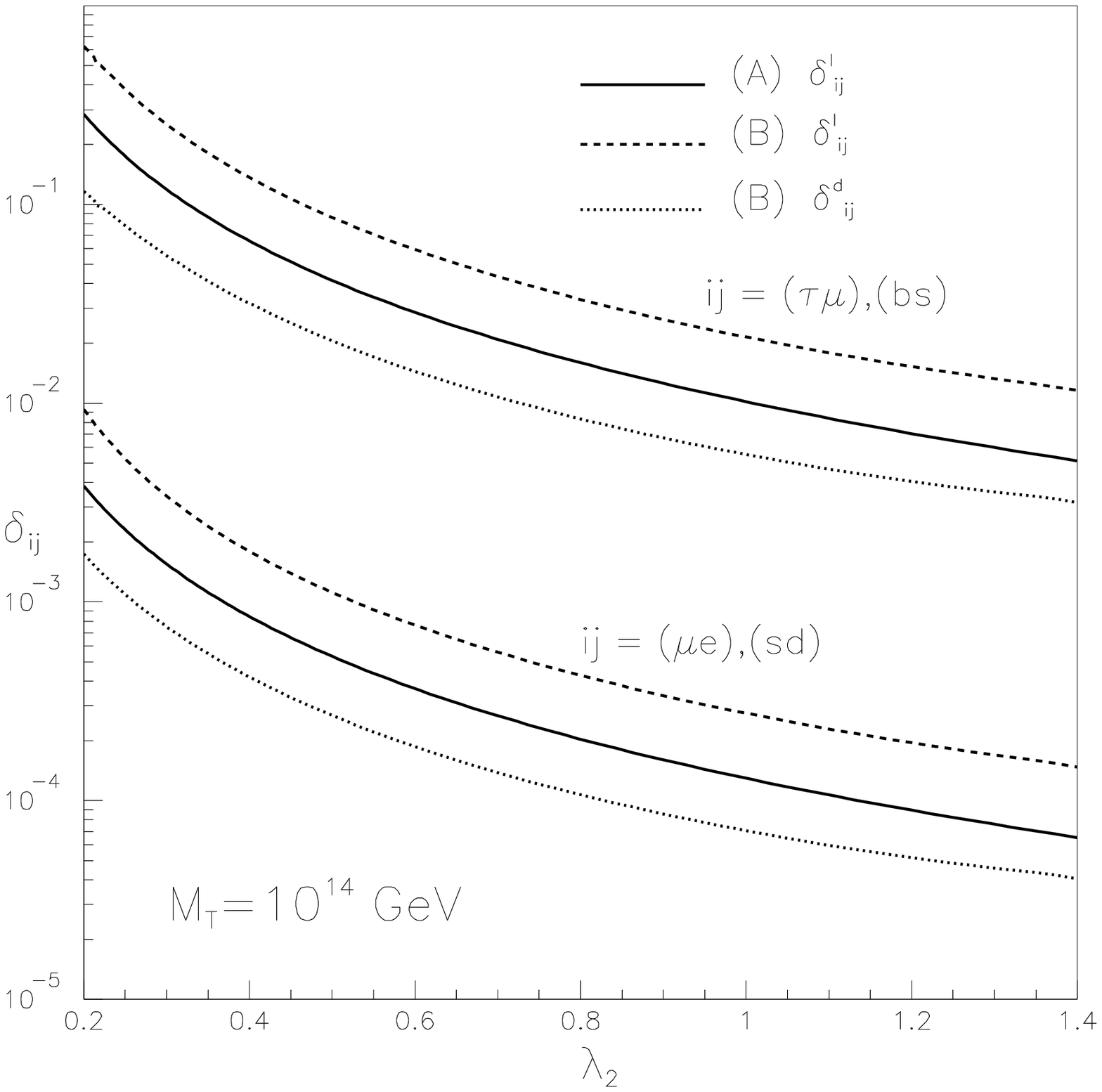,height=9.cm,width= 8.6cm}
\vglue 0.9cm
\hglue -0.8cm
\epsfig{file=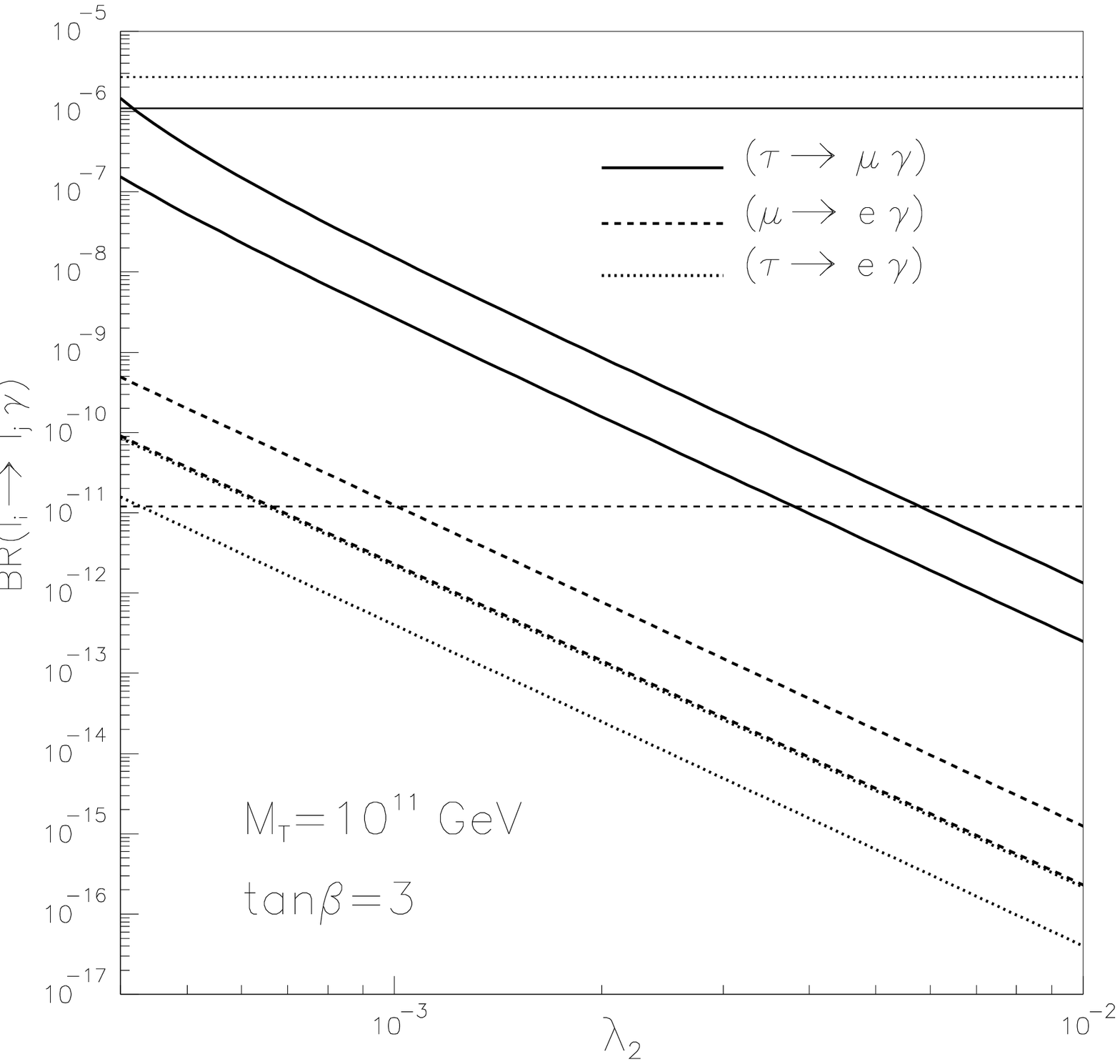,height=8.5cm,width= 8.6cm}
\vglue -8.5cm
\hglue 7.6cm 
\epsfig{file=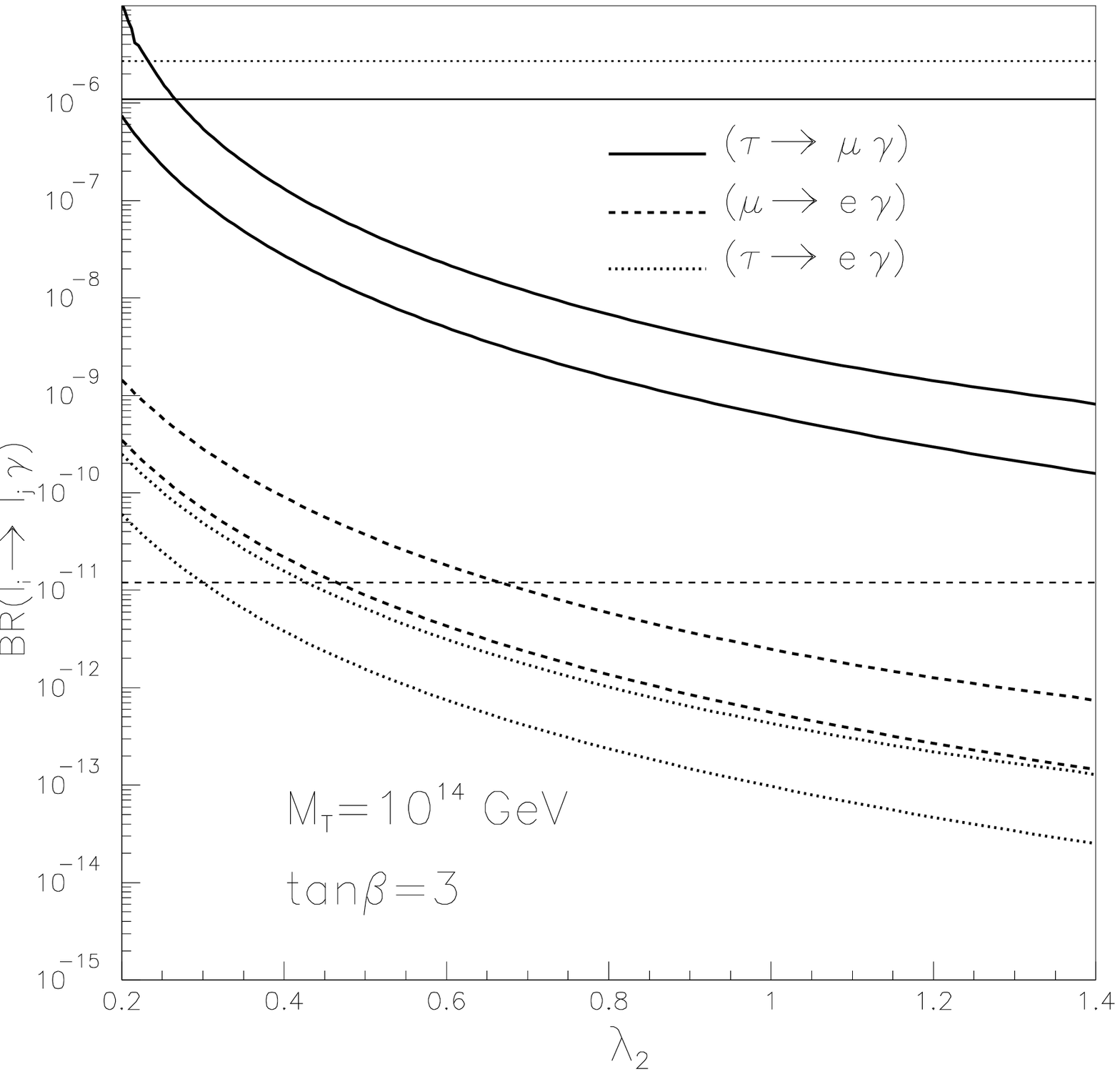,height=8.5cm,width=8.6cm}
\caption{\small 
In the upper panels we show the flavour-violation parameters 
$\delta^\tl$ and $\delta^\td$ as a function of the coupling constant 
$\la_2$ (at $M_T$) for $M_T= 10^{11}~{\rm GeV}$ (left)  and 
$M_T= 10^{14}~{\rm GeV}$ (right) in both scenarios (A) and (B).
As for the neutrino parameters, we have taken $\theta_{23}=45^\circ, 
\theta_{12}=33^\circ, \theta_{13}= 0$ and 
$m_1=0, m_2= 7 \times 10^{-3}~{\rm eV}, 
m_3= 5\times 10^{-2}~{\rm eV}$.   
In correspondence of each scenario, the upper 
lines refer to $\delta^\tl_{\tau \mu}$ and $\delta^\td_{bs}$ 
and the lower ones to 
$\delta^\tl_{\mu e}$ and $\delta^\td_{sd}$ (the $\delta^\td$ are 
non-zero only in the case (B)). We recall  that 
$\delta^\tl_{\tau e}=\delta^\tl_{\mu e} $ as well as 
$\delta^\td_{bd}=\delta^\td_{sd}$. 
In the lower panels we show the resulting branching-ratios 
for the decays $\tau\to \mu \ga, \mu\to e \ga$ and $\tau\to e \ga$. 
The horizontal lines denote the experimental upper bounds of such 
BRs.  For each BR the lower  and upper  curve is obtained 
in the scenario (A) and (B), respectively. 
In all panels, we have fixed $A_0=m_0= 200~$GeV at $M_G$; the 
corresponding average slepton mass at low energy is 
$m_\tl \sim 300~$GeV (250~GeV) for $M_T= 10^{11}~{\rm GeV}$ 
($10^{14}~{\rm GeV}$). Finally,  
the parameters  fixed at low energies are  $\tan\beta=3$, $\mu =300~$
GeV and 
$M_2=180~$GeV; the latter corresponds 
to $M_g = 490~{\rm GeV}$ ($290~{\rm GeV}$) at $M_G$ for 
 $M_T= 10^{11}~{\rm GeV}$  ($10^{14}~{\rm GeV}$).
We have also set $\la_1 =\la_2$ at $M_T$.
}
\label{f2}
\vskip -0.1cm
\end{figure}

\begin{figure}[p]
\vskip -2.cm
\hglue -0.8cm
\epsfig{file=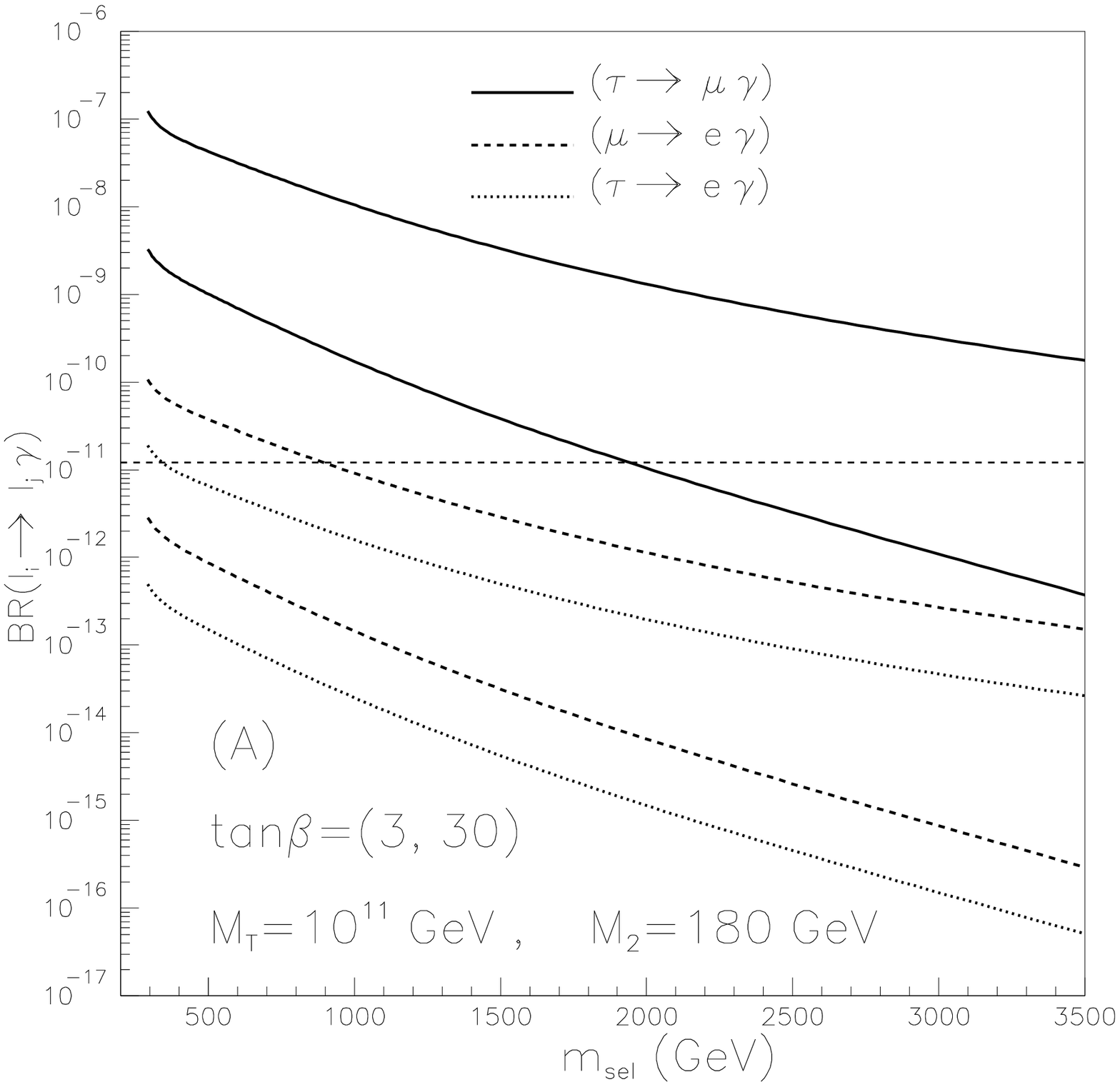,height=9.cm,width= 8.6cm}
\vglue -9.cm
\hglue 7.6cm
\epsfig{file=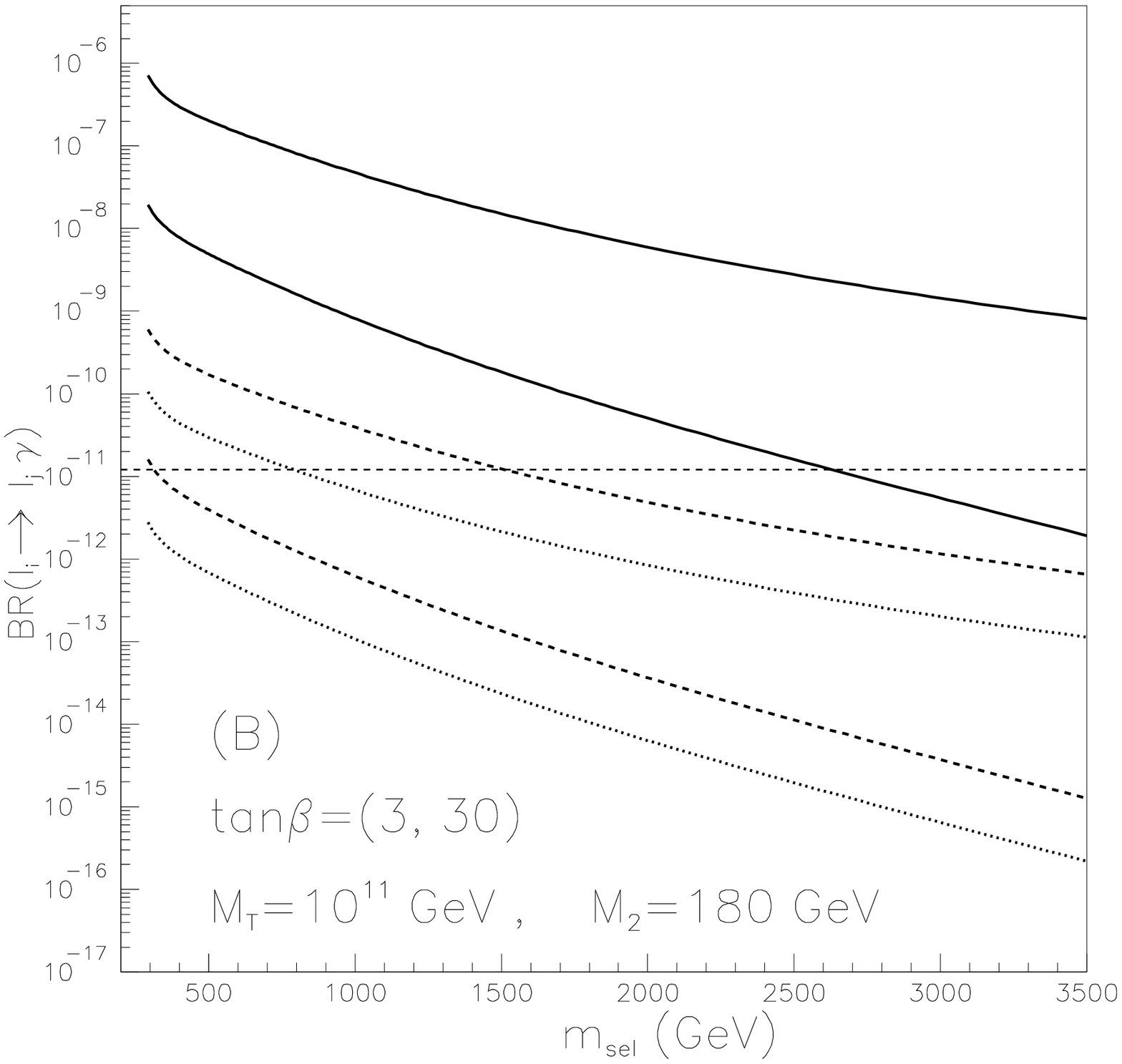,height=9.cm,width= 8.6cm}
\vglue 0.9cm
\hglue 3cm
\epsfig{file=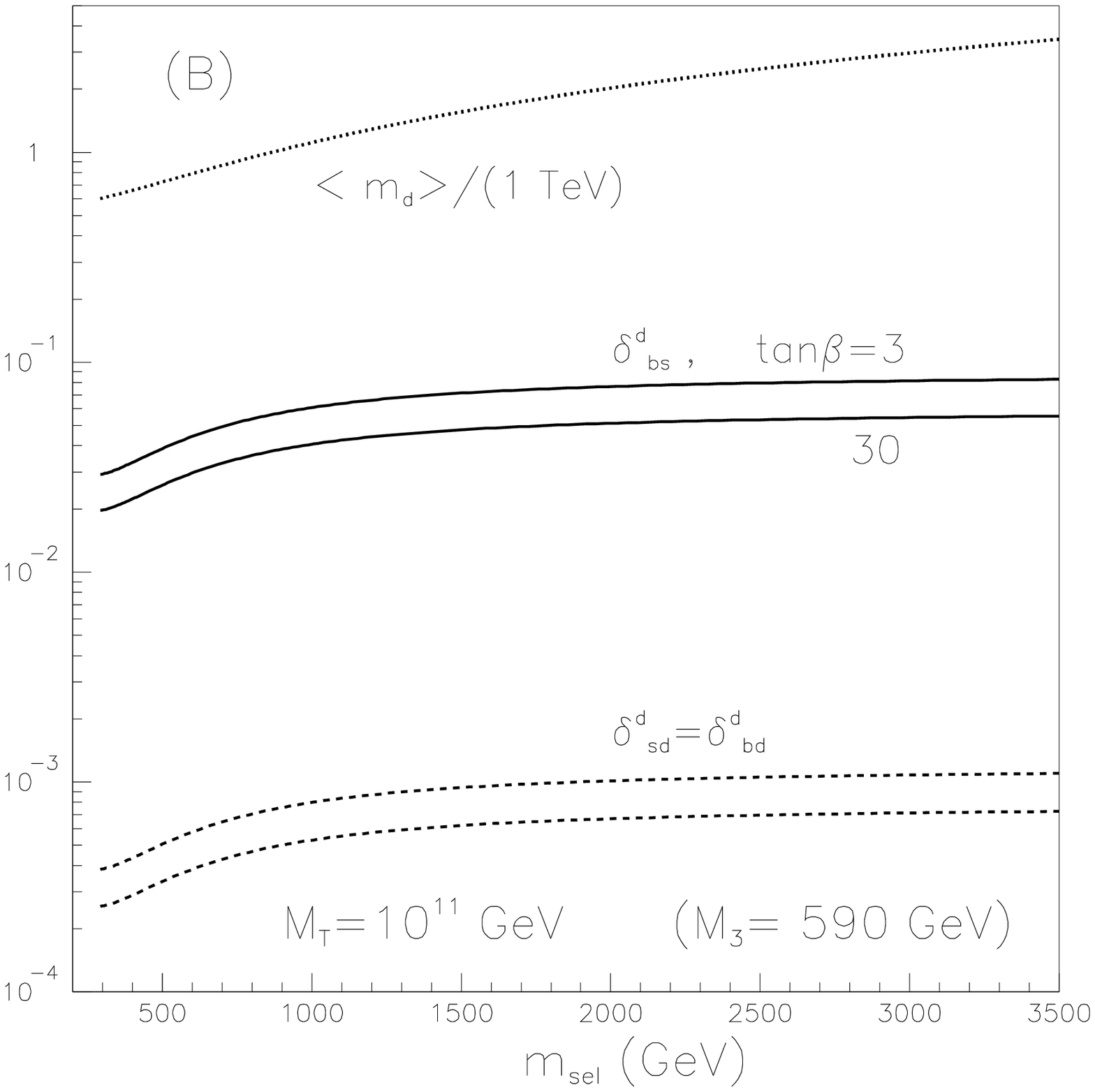,height=8.5cm,width= 8.6cm}
\caption{\small   
In the upper panels we display the behaviour of the 
$\ell_i \to \ell_j \ga$ BRs as a function of the `left-handed' 
selectron mass in both scenarios (A) (left panel) and (B) (right 
panel) for $M_T= 10^{11} ~{\rm GeV}$ and $\la_2 =10^{-3}$ (at $M_T$). 
For each BR the lower and upper line refers to the case with 
$\tan\beta =3$ and 30, respectively. 
The horizontal dashed-line marks the $BR(\mu\to e\ga)$ 
experimental bound.
The gaugino mass and $\mu$  have been taken as in Fig.~\ref{f2}. 
In the lower panel 
we show the corresponding parameters $\delta^\td_{ij}$ 
as emerge in the scenario (B), as a function 
of the selectron mass (solid and dashed lines). 
The dotted lines denotes the average $m_\td$ mass. 
The value of the gluino mass $M_3$ as obtained at low energy 
is also indicated.  
}
\label{f3}
\vskip -0.1cm
\end{figure}
%
The main feature of our picture is  the possibility to 
relate in a {\it  model-independent} way 
the LFV of different sectors, 
as  eqs.~(\ref{LF23-12}) and (\ref{LF13-12}) demonstrate.
Therefore, if we take the corresponding ratio of the 
BRs and  take into account the estimate in (\ref{predi}), we find:
\be{ratepredi1}
\frac{BR(\tau \to \mu\ga)}{BR(\mu \to e\ga)} \approx 
\left(\frac{(\bm^2_{\tl})_{ \tau \mu}}{(\bm^2_{\tl})_{\mu e}}\right)^2 
\frac{BR(\tau \to \mu \nu_\tau \bar{\nu}_\mu)}
{BR(\mu \to e \nu_\mu \bar{\nu}_e)} \sim 10^3  . 
\ee
For the case of SMA solution, this would become much larger, i.e. 
$\sim 10^7$.
Analogously we can predict 
\be{ratepredi2}
\frac{BR(\tau \to e \ga)}{BR(\mu \to e\ga)} \approx 
\left(\frac{(\bm^2_{\tl})_{ \tau e}}{(\bm^2_{\tl})_{  \mu e}}\right)^2 
\frac{BR(\tau \to e \nu_\tau \bar{\nu}_e)}
{BR(\mu \to e \nu_\mu \bar{\nu}_e)} \sim 10^{-1}  . 
\ee 

Let us now discuss the results obtained 
by a more detailed numerical study 
with exact solutions of the RGEs and exact diagonalization 
of matrices involved in the computation of the branching ratios. 
First in Fig.~\ref{f2}  we present 
the behaviour of the LFV 
parameters $\delta^\tl_{ij}$ and $\delta^\td_{ij}$ defined as
\be{deltas}
\delta^\tl_{ij} = \frac{|(\bm^2_\tl)_{ij}|}{m^2_\tl} , ~~~~~~~
\delta^\td_{ij} = \frac{|(\bm^2_\td)_{ij}|}{m^2_\td} , 
\ee
($m^2_\tl, m^2_\td$ are the average $\tl$ and $\td$ squared masses) 
as a function of the coupling $\la_2$ in both scenarios (A) and (B), for 
two values of $M_T$, i.e. 
$M_T= 10^{11} ~{\rm GeV}$ (upper 
left  panel) and 
$M_T= 10^{14} ~{\rm GeV}$ (upper right panel) and for representative 
values of other parameters. For each 
scenario the upper and lower curve refers to $\delta^\tl_{\tau \mu}$ 
and    $\delta^\tl_{\mu e}$ (or $\delta^\tl_{\tau e}$, 
cfr. eq.~(\ref{predi31})), respectively,   
and similarly   for $\delta^\td_{bs}$  and $\delta^\td_{sd}$, in the
scenario (B). 
Due to the quadratic  dependence in the RGEs   
on the LFV Yukawa matrices $\bY_T, \bY_S, \bY_Z$,  
the $\delta$ parameters scale 
as $(\la_2)^{-2}$. We notice that the size of LFV in the 2-3 and 1-2  
sectors maintains a constant ratio, $\sim 10^2$,
independently of the scale $M_T$ (cfr. the estimate 
in eq.~(\ref{predi})). 
In correspondence of each value of the scale $M_T$, there is a minimum value 
for  $\la_2$, below which the RGE solutions blow up\footnote{
One has also to check that $\lambda_2$ is not too large at $M_T$ so that it 
remains  perturbative up to $M_G$ (cfr. the $\la_2$-RGE in 
eq.~(\ref{rge-ye}) in Appendix).}, 
which is  approximately 
$\la^{\rm min}_{2 } \sim  3 \times 10^{-4} 
(M_T/ 10^{11} ~{\rm GeV})$.
We can notice that in the scenario (B)  $\delta^\tl_{ij}$ gets 
larger than in case (A) by about a factor 2. Moreover,  
$\delta^\td_{ij}$  are about a factor 2 smaller than $\delta^\tl_{ij}$. 
\begin{figure}[p]
\vskip -2.cm
\hglue -0.7cm
\epsfig{file=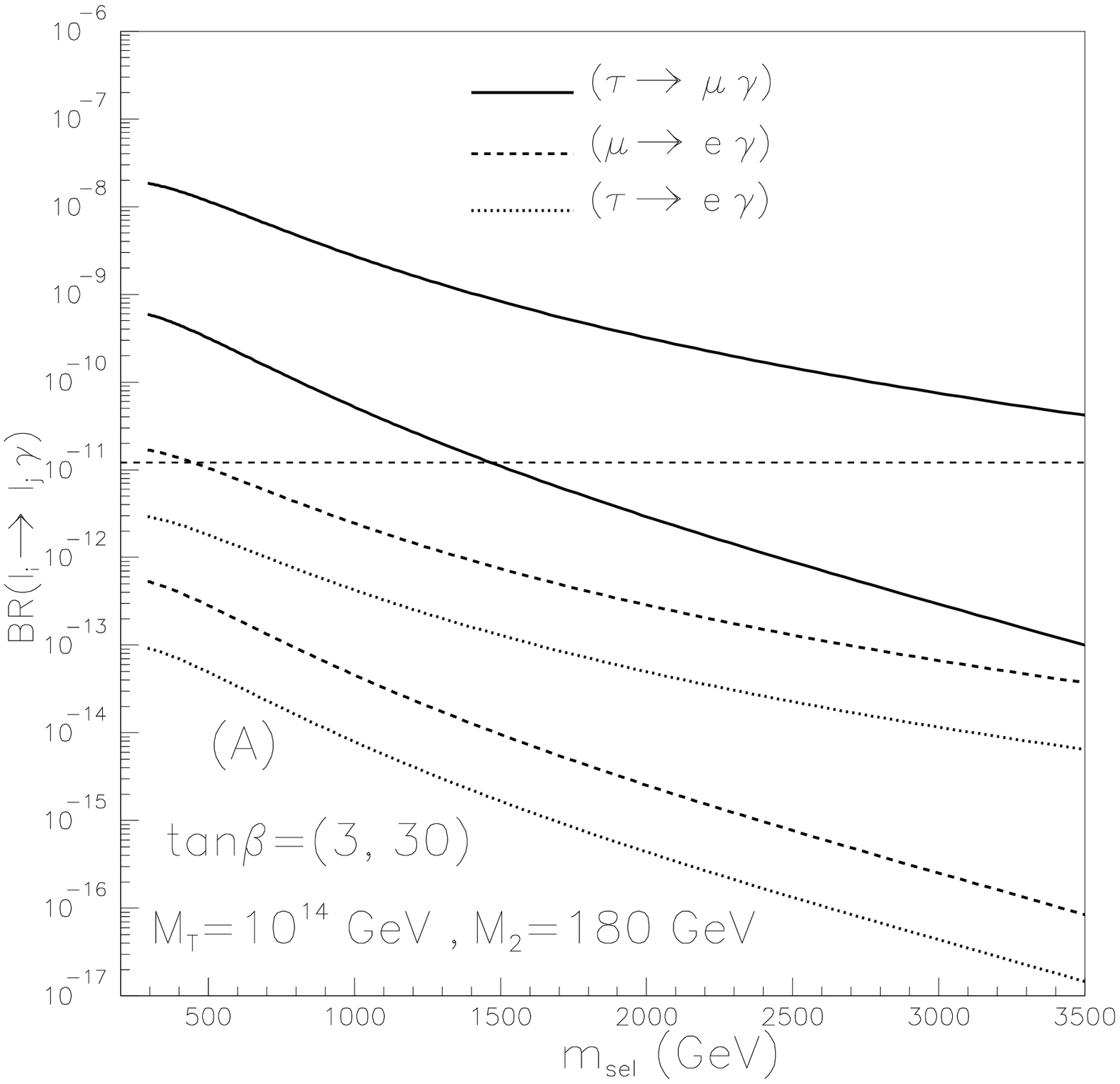,height=7.8cm,width= 8.6cm} 
\vglue -7.8cm
\hglue 7.5cm  
\epsfig{file=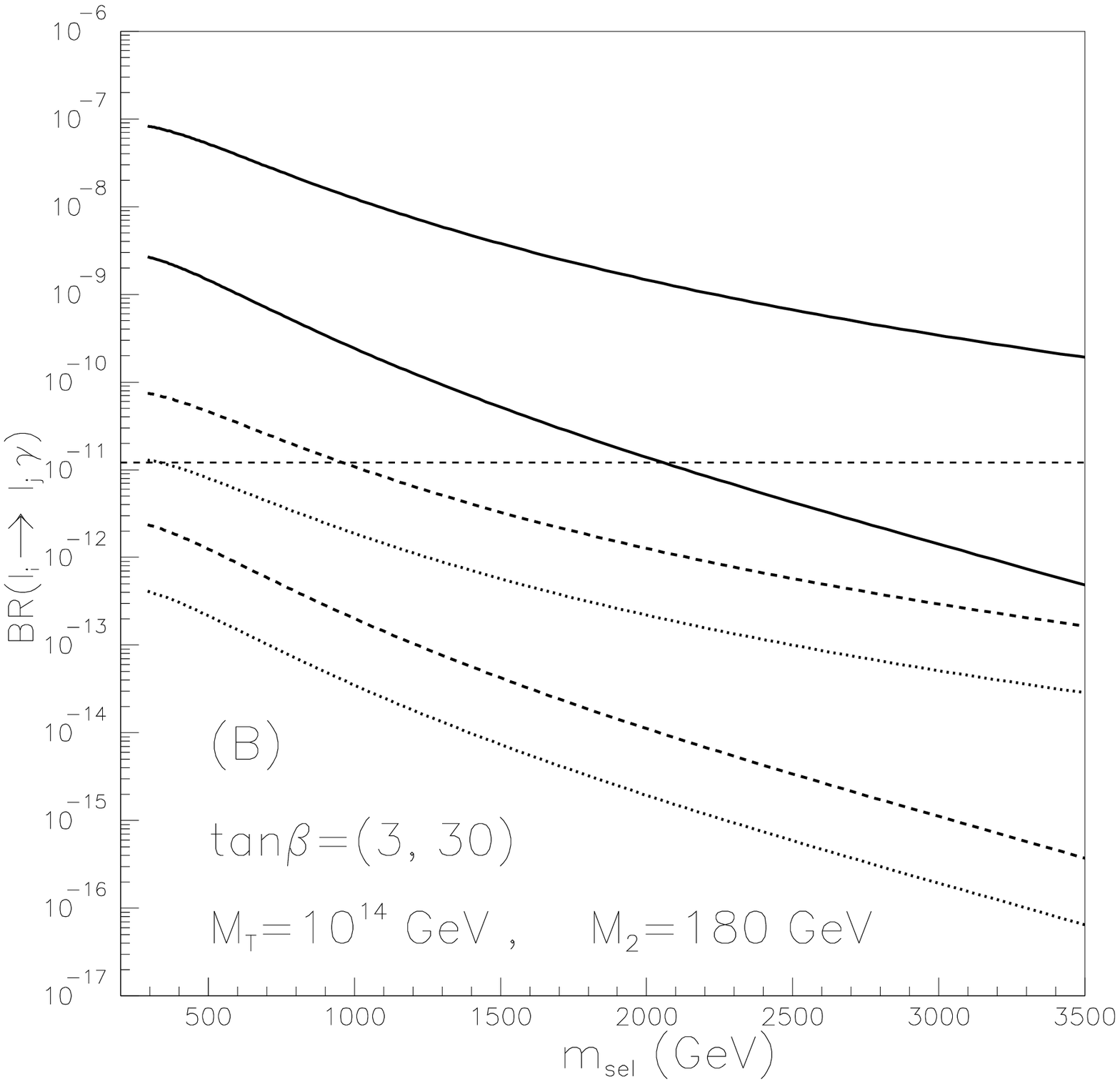,height=7.8cm,width= 8.6cm}
\vglue 0.9cm
\hglue 3cm
\epsfig{file=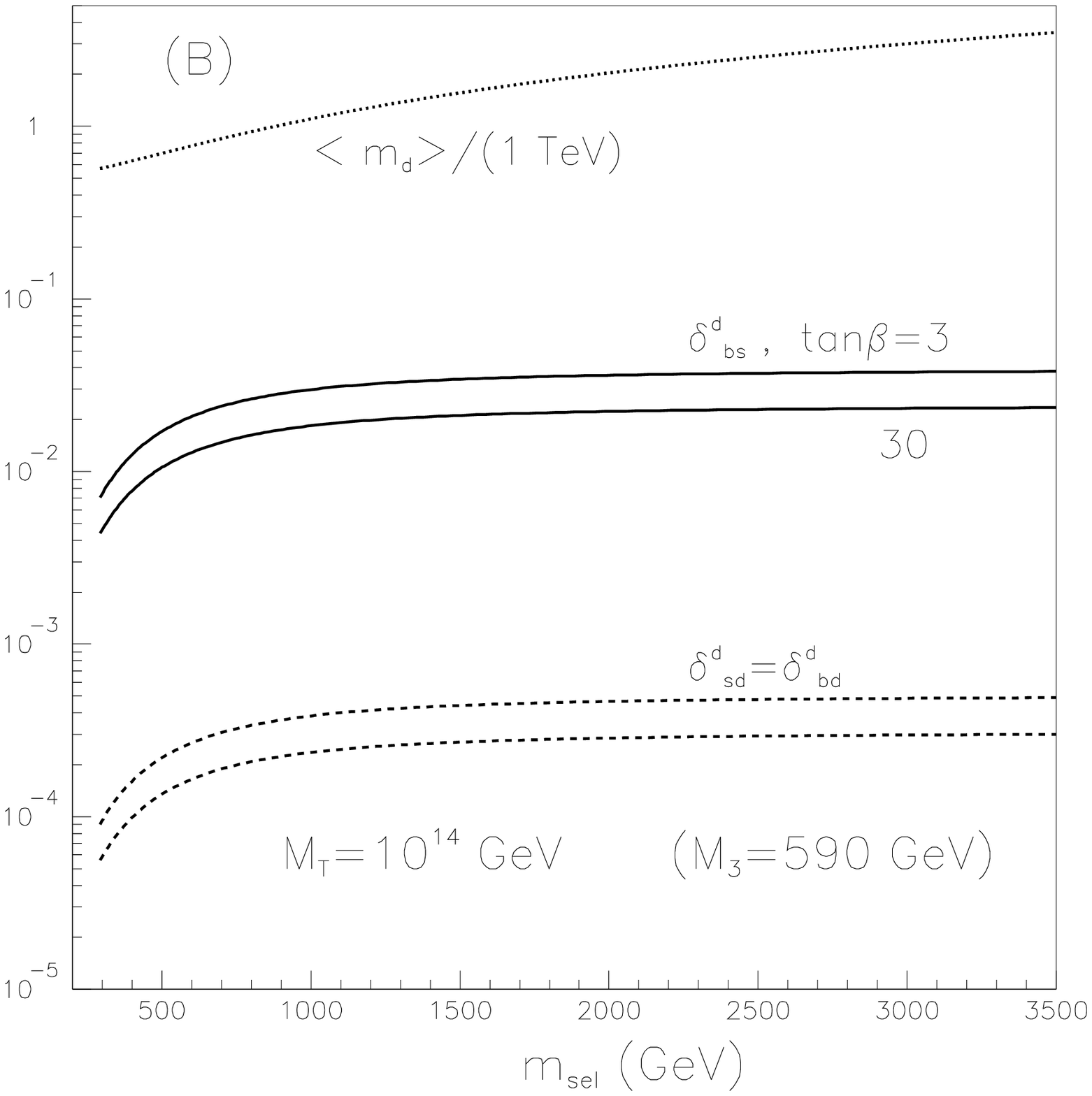,height=8.5cm,width= 8.6cm}
\caption{\small 
The same as in Fig.~\ref{f3} for $M_T= 
10^{14} ~{\rm GeV}$ and $\la_2=1$.
}
\label{f4}
\end{figure}
This is due to the fact that in the evolution to low-energy 
the squark mass gets heavier than the slepton mass, because of the 
gluino-driving. This different 
increase is reflected mostly in the average scalar masses 
(which are determined at the SUSY scale) and to a less extent 
in the off-dagonal entries $(\bm^2_\tl)_{ij},  (\bm^2_\td)_{ij}$ 
(which instead are determined at the intermediate scale $M_T$). 
The partial compensation of these effects explains why, 
for $M_T=10^{11}~$GeV, for instance,   
the ratio $\delta^\tl/\delta^\td$ is smaller than 4, even though  
$m_\tl\sim 300~$GeV and $m_\td\sim 600$~GeV. 

In the same figure we also show the 
behaviour of the  branching ratios (lower panels) 
for $\tan\beta =3$, 
as a representative case. 
For each decay the lower and  upper curve 
refers to the scenario (A) and (B), respectively.
It is striking to notice that the 
constant-ratio rule displayed in eqs.~(\ref{ratepredi1}- 
\ref{ratepredi2}) between the $BR$ of the decays $\tau\to \mu \ga$  
or $\tau\to e\ga$ 
with that of  $\mu \to e \ga$ is preserved in each case (A) and (B) 
as well as for any value of $M_T$,  irrespectively 
of the value of $\la_2$. Moreover, in the scenario (B) with all 
the Yukawa couplings $\bY_{T,S,Z}$ at work 
these rates are  larger since 
the LFV in the slepton masses is enhanced, as seen in the upper panels. 
The present bound on the $BR(\mu\to e\ga)$ (shown by dashed horizontal line) 
constrains  
$\la_2$ to be larger than $\sim 7 \times 10^{-4}$ and $\sim 0.46$ for $M_T=
10^{11}~{\rm GeV}$ and $10^{14}~{\rm GeV}$,  respectively in the 
case (A). In the picture (B) these bounds get a bit more stringent, 
 $ \la_2 \gsim 2 \times 10^{-3}$ and $ \la_2 \gsim 0.65$    
  for $M_T=
10^{11}~{\rm GeV}$ and $10^{14}~{\rm GeV}$,  respectively.
The $T$-induced seesaw strict predictions for 
the ratio of the $BR$s tell us that, 
in view of the future experimental sensitivity, 
both  decays  $\tau\to \mu \ga$ and $\mu \to e\ga$ could be observed   
if the latter has at least a branching ratio of order $10^{-12}$. For 
$BR(\mu\to e\ga) < 10^{-12}$, we have that 
$BR(\tau \to \mu \ga) < 10^{-9}$, below the expected sensitivity. 
On the other hand, if we depart from the limit $\theta_{13}=0$ 
and allow  $\theta_{13}$  as large as 0.1, 
we would get  $BR(\tau \to \mu \ga)\sim 10\cdot BR(\mu \to e \ga)$. 
As for  $BR(\tau \to e \ga)$ it is always predicted to be one order 
of magnitude smaller than $BR(\mu\to e\ga)$.

In Fig.~\ref{f3} we show $BR(\mu\to e \ga)$, $BR(\tau \to \mu  \ga)$ 
and $BR(\tau \to e \ga)$ 
as a function of the left-handed selectron mass at low-energy with 
$M_T=10^{11} ~{\rm GeV}$ and $\la_2=10^{-3}$ 
for both  scenarios (A) (upper left panel) 
and (B) (upper right panel).  For each  $BR$, the upper and lower 
curves correspond to $\tan\beta =30$ and 3, respectively. 
In the scenario (A) 
for $\tan\beta =3$ there are no constraints 
on $m_{\tilde{e}}$ from the $BR(\mu \to e \ga)$ bound, while for 
$\tan\beta =30$ we have $m_{\tilde{e}}\gsim 900~$ GeV. More stringent 
lower bounds on $m_{\tilde{e}}$ can be deduced in the scenario (B). 
Therefore in the allowed range for $m_{\tilde{e}}$, 
we have $BR(\tau \to \mu  \ga)\lsim 2 \times 10^{-8}$ and 
$BR(\tau \to e  \ga)\lsim 2 \times 10^{-12}$. 
Again notice  that 
the  constant-ratio rule for the $BR$s is maintained also in this case 
where $m_\tel$ is varied, confirming the fact that this rule 
depends only on the low-energy neutrino parameters and not on 
the details of the model, such as the soft-breaking parameters or 
$M_T$. In the lower panel, we show the behaviour of $\delta^\td$ 
and the average squark mass $m_\td$ versus $m_\tel$. 
The fact that $\delta^\td$ are about a factor 1.5 larger for 
$\tan\beta =3$ as compared to the case with $\tan\beta =30$ 
is due to the combined increase for lower values of $\tan\beta$ 
of both $\bY_\nu$ (see eq.~\ref{T-mass}))  
and the top Yukawa coupling, 
$Y_t \sim m_t/(v \sin\beta) $, 
which influences the running of $\bY_\nu$ 
from low-energy up to $M_T$ (see eq.~(\ref{rge5}) in Appendix).
We recall that the whole increase of $\bY_\nu$ implies an increase of 
$\bY_T$ as well of $\bY_Z$ and $\bY_S$ through the relation (\ref{unif1}). 
Then the quantities   $\bY_Z\bY^\dagger_Z$ and 
$\bY_S\bY^\dagger_S$ (scaling as $1/\sin^4\beta$) trigger  
the flavour-violation in the matrix $\bm_\td$.
The information on $\delta^\td$ and  $m_\td$ could be useful for the 
comparison with the present bounds on the flavour-violation 
parameters $\delta^\td$ 
extracted from the meson mixing measurements \cite{masiero}. 
To this purpose we need to know also that 
the gluino mass at low-energy is $M_{3}\sim  600~{\rm GeV}$ 
for $M_T=10^{11} ~{\rm GeV}$. For example, for $\tan\beta =3$ 
and $m_\tel > 300~$ GeV, we have $m_\td \sim 600~$GeV  
and $\delta^\td_{sd} (~{\rm or} ~\delta^\td_{bd}) 
 \approx 4\times 10^{-4}$ which is below  
the limits from the $K^0-\bar{K}^0$ (or $B_d-\bar{B}_d$) 
mixing parameter.

In Fig.~\ref{f4} the same analysis has been performed 
for $M_T= 10^{14} ~{\rm GeV}$ and we have chosen $\la_2=1$, 
in such a way 
that the ratio $M_T/\la_2$ is the same as in the previous example.
In this way, the size of the matrix $\bY_T$ is the same at the 
scale $M_T$.
All other parameters, 
such as $M_2$ at low energy, are the same as in Fig.~\ref{f3}. 
Upon comparing with the previous  case with lower $M_T$,  
we observe that for larger $M_T$ 
the BRs are smaller by a factor 5 which is due to the 
smaller energy interval of the running, namely 
$\left[\log (M_G/10^{11}~{\rm GeV})/\log (M_G/10^{14}~{\rm GeV})\right]^2 
\approx 5$. For the same reason, for $M_T= 10^{14} ~{\rm GeV}$
the parameters $\delta^\td$ are smaller by a factor 2 or so.

\section{Conclusions}
The neutrino experimental observations pointing to sizeable 
lepton mixing have been encouraging  to further investigate the implications 
of lepton-flavour violation in extensions of the Standard Model. 
This work may be placed among these attempts. In particular, 
we have considered the SUSY seesaw mechanism obtained 
through the exchange of heavy $SU(2)$ triplets. On comparing  with the 
more popular seesaw scenario with the exchange of heavy singlets, 
our scenario is more predictive since the source of LFV at high energy, 
i.e. the Yukawa matrix $\bY_T$, can be directly connected to 
the low-energy observables, encoded in the coupling matrix $\bY_\nu$. 
The Yukawa $\bY_T$ induces radiative corrections in the 
mass matrix of the sleptons $\tl$ and as a result, even in the case 
of universal scalar masses at the GUT scale, LFV off-diagonal entries 
are generated. Therefore the flavour structure  of the slepton mass 
matrix can be determined solely in terms of the low-energy neutrino 
parameters, i.e. the neutrino masses and mixing angles. 
The most remarkable feature of this scenario is that there is 
 a rigid entanglement  of the 
flavour-violation among different generations, as displayed in 
eqs.~(\ref{predi}-\ref{predi31}),  which does not depend on other 
details of the theoretical  framework. This implies in particular 
that  the ratio of the branching ratios of the decay 
$\tau \to \mu \ga$ (or $\tau \to e \ga$) 
and $\mu \to e \ga$  can be predicted and turns out 
to be $\sim 10^3$ (or ~$10^{-1}$). 
We have first derived 
these estimates by only taking into   account the neutrino 
parameters (\ref{ratepredi1}-\ref{ratepredi2}) and  have confirmed them
 by more detailed numerical computations, as shown in Figs.~2, 3 and 4. 
Furthermore we have embedded this picture in a `minimal' $SU(5)$ scenario 
in which the triplet states $T$ fill the 15-representation together 
with other coloured partners $S, Z$. In such a case flavour violation 
is also induced by radiative corrections on the mass matrix of 
the $d^c$-squarks.  
By imposing the GUT scale $SU(5)$-universality relation among 
the Yukawa couplings of those states to the matter multiplets, 
we find that the size of flavour violation in the lepton and quark sectors  
is comparable. This implies that, similarly 
to what happens in the lepton sector, the amount of 
flavour-violation between different quark sectors is 
strongly correlated. 
For example, one could predict the supersymmetric 
contribution to  $B_{s}-\bar{B_s}$  
mixing in terms of that to $K^0-\bar{K^0}$ mixing. 
It would be interesting to further explore this point and other 
implications of the SUSY $T$-induced seesaw.

 \vspace{0.5cm}

{\bf Acknowledgments} 
This work  was   partially supported
by the European Union under the contracts 
HPRN-CT-2000-00148 (Across the Energy Frontier) and
 HPRN-CT-2000-00149 (Collider Physics).

\vspace{0.8cm}
\noindent
{\bf \Large Appendix}
\vspace{0.4cm}

\noindent
In this appendix we first present the parametrization used for the chargino 
and neutralino mass matrix since this determines the relative sign 
between the Yukawa and gauge terms\footnote{
We have indeed found some discrepancy 
in the literature. Our results, for example, are in agreement 
with the RGEs of 
\cite{falck}. In the latter work 
there is consistency between the 
sign of the gaugino mass terms and the RGEs of the trilinear terms 
provided  a (missed)  minus sign is accounted  
in front of the gaugino mass in the matrix of eq.~(2.7) (or, equivalently, 
provided the $i$ factor in the off-diagonal blocks is removed).}  
in the 
renormalization group equations of the trilinear soft-breaking terms 
(see below (\ref{soft-A})). 
Then we have determined the renormalization group equations in the 
MSSM with the $15 +\ov{15}$ representation of $SU(5)$ at 
one-loop, which are therefore valid in the energy range between 
$M_G$ and the triplet mass scale $M_T$.

The chargino mass matrix term is given by:
\be{mch}
-{\cal L}_{\rm ch}=
\begin{array}{ccc}
 & 
&\\ \vspace{2mm}
\left(\begin{array}{c}
 {\tilde{W}^{+}}\\  {\tilde{H}^+}_{2 } 
\end{array}\right)^T ~
\!\!\!\!\!\!\!\!\!&{\left(\begin{array}{cc}
M_2 & g v_2 \\ 
g v_1  & \mu \end{array}\right)} &
 \! \!\left(\begin{array}{c}
 {\tilde{W}^{-}}\\  {\tilde{H}^-}_{2 } 
\end{array}\right)
\end{array}\! +{\rm h.c.} , 
\ee
and that regarding the  neutralino mass sector is:
\be{neu}
-{\cal L}_{\rm n}= 
\frac12 \!\!
\begin{array}{ccccc}
 & 
& \\ 
\left(\begin{array}{c}
\tilde{B}\\ \,\,\,\, \tilde{W}^0 \\  \tilde{H}^0_{1}   \\  \tilde{H}^0_{2} 
\end{array}\right)^T\!\!\!\!\!&{\left(\begin{array}{cccc}
M_1 & 0  & - \frac{1}{\sqrt2} g'v_1   & \frac{1}{\sqrt2} g'v_2  \\ 
0 & M_2 &  \frac{1}{\sqrt2} g v_1 & - \frac{1}{\sqrt2} g v_2 \\ 
- \frac{1}{\sqrt2} g'v_1 & \frac{1}{\sqrt2} gv_1 & 0 &-\mu \\  
\frac{1}{\sqrt2} g'v_2 & -\frac{1}{\sqrt2} gv_2 & -\mu 
& 0 
\end{array}\right)} & 
\!\! \left(\begin{array}{c}
\tilde{B}\\ \,\,\,\, \tilde{W}^0 \\  \tilde{H}^0_{1}   \\  \tilde{H}^0_{2} 
\end{array}\right)
\end{array}  ~ +{\rm h.c.} .
\ee   

The renormalization group equations for the gaugino masses $M_a,~(a=1,2,3)$ 
are:
\be{grge}
16\pi^2 \frac{d M_a}{dt} = 2 g^2_a B_a M_a , 
\ee 
where the coefficients $B_a$ are given in eq.~(\ref{betas}).  
The RGEs for the Yukawa couplings are: 
\beqn{rge-ye}
16\pi^2 \frac{d \bY_T}{dt}\!\!\!&=\!\!\!& 
\bY_T\left[ -\frac95 g^2_1 - 7 g^2_2 +
 \bY^\dagger_e \bY_e + 6\bY^\dagger_T\bY_T +{\rm Tr}
(\bY^\dagger_T \bY_T) +\bY^\dagger_Z \bY_Z +|\la_1|^2\right]\nonumber \\
&&+(\bY^T_e \bY^\star_e+ \bY^T_Z \bY^\star_Z) \bY_T ,  \nonumber \\
16\pi^2 \frac{d \bY_S}{dt}\!\!\!&=\!\!\!& 
\bY_S\left[ -\frac45 g^2_1 - 12 g^2_3 +
 2\bY^\star_d \bY^T_d + 8\bY^\dagger_S\bY_S +{\rm Tr}
(\bY^\dagger_S \bY_S) +2 \bY^\star_Z \bY^T_Z \right]\nonumber \\
&&+2 (\bY_d \bY^\dagger_d + \bY_Z \bY^\dagger_Z) \bY_S ,  \nonumber \\
16\pi^2 \frac{d \bY_Z}{dt}\!\!\!&=\!\!\!& 
\bY_Z\left[ -\frac{7}{15} g^2_1 - 3 g^2_3 --\frac{16}{3} g^2_3+
 \bY^\dagger_e \bY_e + 5\bY^\dagger_Z\bY_Z +{\rm Tr}
(\bY^\dagger_Z \bY_Z) +3 \bY^\dagger_T \bY_T \right]\nonumber \\
&&+2 (\bY_d \bY^\dagger_d + 2\bY_S \bY^\dagger_S) \bY_Z ,  \nonumber \\
16\pi^2 \frac{d \bY_e}{dt} \!\!\!&=\!\!\!& \bY_e \left[
 -\frac{27}{15} g^2_1 - 3 g^2_2  
+ 3(\bY^\dagger_e \bY_e + \bY^\dagger_T\bY_T +\bY^\dagger_Z\bY_Z+ |\la_1|^2) +
 {\rm Tr}(\bY^\dagger_e \bY_e +3 \bY^\dagger_d \bY_d)\right] , \nonumber \\ 
16\pi^2 \frac{d \bY_d}{dt}\!\!\!& = \!\!\!&\bY_d \left[
 -\frac{7}{15} g^2_1  - 3 g^2_2  -\frac{16}{3} g^2_3   
+ 3(\bY^\dagger_d \bY_d + |\la_1|^2 ) +
\bY^\dagger_u\bY_u \right.\nonumber \\
&& \left .
+ {\rm Tr}(\bY^\dagger_e \bY_e +3 \bY^\dagger_d \bY_d) \right] 
+2(\bY_Z \bY^\dagger_Z +2 \bY_S \bY^\dagger_S)\bY_d  , \nonumber \\ 
16\pi^2 \frac{d \bY_u}{dt}\!\!\! &=\!\!\! &\bY_u \left[
 -\frac{13}{15} g^2_1 - 3 g^2_2    -\frac{16}{3} g^2_3   
+ 3\bY^\dagger_u \bY_u + \bY^\dagger_d\bY_d + 3|\la_2|^2 +
 3{\rm Tr}(\bY^\dagger_u \bY_u) \right] , \nonumber \\ 
16\pi^2 \frac{d \la_1} {dt}\!\!\! &=\!\!\! & \la_1 \left[7 \la^2_1 + {\rm Tr}(
\bY_T \bY^\dagger_T+ 2\bY_e  \bY^\dagger_e +6\bY_d \bY^\dagger_d) 
-\frac95 g^2_1-7g^2_2 \right] ,  \nonumber\\
16\pi^2 \frac{d \la_2} {dt}\!\!\! &=\!\!\! & \la_2 \left[7 \la^2_2 + 6{\rm Tr}
(\bY_u \bY^\dagger_u)
-\frac95 g^2_1-7g^2_2 \right] ,  
\eeqn  
where $\bY_d$ and $\bY_u$ are the Yukawa coupling matrix of the down and up 
quarks, respectively.
Regarding the  mass parameters of the superpotential, the 
RGEs are:
\beqn{mulike}
16\pi^2 \frac{d \mu} {dt}\!\!\! &=\!\!\! & \mu \left[3 {\rm Tr}(
\bY^\dagger_u \bY_u+ \bY^\dagger_d \bY_d) + 
{\rm Tr}
(\bY^\dagger_e \bY_e) +3(|\la_1|^2+|\la_2|^2 ) 
-\frac35 g^2_1-3g^2_2 \right] ,  \nonumber\\
16\pi^2 \frac{d M_T} {dt}\!\!\! &=\!\!\! & M_T \left[{\rm Tr}
(\bY^\dagger_T \bY_T)
+ |\la_1|^2+|\la_2|^2  
-\frac35 g^2_1-8g^2_2 \right] ,  \nonumber\\
16\pi^2 \frac{d M_S} {dt}\!\!\! &=\!\!\! & M_S \left[ {\rm Tr}
(\bY^\dagger_S \bY_S)
-\frac{16}{15} g^2_1- \frac{40}{3}g^2_3 \right] ,  \nonumber\\
16\pi^2 \frac{d M_Z} {dt}\!\!\! &=\!\!\! & M_Z \left[ {\rm Tr}
(\bY^\dagger_Z \bY_Z)
-\frac{1}{15} g^2_1 -3g^2_2- \frac{16}{3}g^2_3 \right] .
\eeqn
The RGEs for the  sfermion mass matrices are: 
\beqn{RGEsoft}
16\pi^2 \frac{d \bm^2_{\tl} }{dt} &\!\!\!= \!\!\!& 
\bm^2_{\tl}( \bY^\dagger_e \bY_e
+ 3 \bY^\dagger_T \bY_T+ 3 \bY^\dagger_Z \bY_Z) + 
( \bY^\dagger_e \bY_e + 3 \bY^\dagger_T \bY_T+ 3 \bY^\dagger_Z \bY_Z)
\bm^2_{\tl}  \nonumber \\
&& +2\left(\bY^\dagger_e m^2_{H_1} \bY_e  + 
 \bY^\dagger_e \bm^2_{\te} \bY_e +
 3\bY^\dagger_T \bm^{2 T}_{\tl} \bY_T + 3\bY^\dagger_Z \bm^{2 }_{\td} \bY_Z + 
 3  \bY^\dagger_T m^{2 }_{T} \bY_T  \right. \nonumber \\
&& \left. +3  
\bY^\dagger_Z m^{2 }_{Z} \bY_Z \right) + 
  2( \bA^\dagger_e \bA_e + 3\bA^\dagger_T \bA_T+3\bA^\dagger_Z \bA_Z)
-\frac65 M^2_1 g^2_1 - 6 M^2_2 g^2_2  , \nonumber \\ 
16\pi^2 \frac{d \bm^2_{\te} }{dt} &\!\!\!= \!\!\!& 
\left(2\bm^2_{\te} + 4 m^2_{H_1}\right) \bY_e\bY^\dagger_e + 4 
\bY_e \bm^2_{\tl}\bY_e^\dagger +2 \bY_e \bY_e^\dagger \bm^2_{\te} +
4\bA_e\bA^\dagger_e -\frac{24}{5} M^2_1 g^2_1 , \nonumber \\
16\pi^2 \frac{d \bm^2_{\td} }{dt} &\!\!\!= \!\!\!& 
2\bm^2_{\td}( \bY_d \bY^\dagger_d + 2\bY_S \bY^\dagger_S +
\bY_Z \bY^\dagger_Z) +2 (
\bY_d \bY^\dagger_d + 2\bY_S \bY^\dagger_S +
\bY_Z \bY^\dagger_Z)
\bm^2_{\td}  \nonumber \\
&& +4\left(\bY_d m^2_{H_1} \bY^\dagger_d  + 
 \bY_d \bm^2_{\tq} \bY^\dagger_d +
 2\bY_S \bm^{2 T}_{\td} \bY_S + \bY_Z \bm^{2 }_{\tl} \bY^\dagger_Z + 
 2  \bY_S m^{2 }_{S} \bY^\dagger_S  \right. \nonumber \\
&& \left. +\bY_Z m^{2 }_{Z} \bY^\dagger_Z \right) + 
 4( \bA_d \bA^\dagger_d + 2\bA_S \bA^\dagger_S+ \bA_Z \bA^\dagger_Z)
-\frac{8}{15} M^2_1 g^2_1 - \frac{32}{3} M^2_3 g^2_3  , \nonumber \\
16\pi^2 \frac{d \bm^2_{\tu} }{dt} &\!\!\!= \!\!\!& 
\left(2\bm^2_{\tu} + 4 m^2_{H_2}\right)\bY_u \bY^\dagger_u  
+ 4 \bY_u \bm^2_{\tq} \bY^\dagger_u +
2\bY_u \bY^\dagger_u\bm^2_\tu+ 4\bA_u \bA^\dagger_u \nonumber \\
&&-\frac{32}{15} M^2_1 g^2_1 - \frac{32}{3} M^2_3 g^2_3 , \nonumber\\
16\pi^2 \frac{d \bm^2_{\tq} }{dt} &\!\!\!= \!\!\!& 
\left(\bm^2_{\tq} + 2 m^2_{H_2}\right)\bY^\dagger_u \bY_u+
\left(\bm^2_{\tq} + 2 m^2_{H_1}\right)\bY^\dagger_d \bY_d
  +\left(\bY^\dagger_u \bY_u+ \bY^\dagger_d \bY_d\right)\bm^2_\tq 
\nonumber \\
&&+ 2 \bY^\dagger_u \bm^2_{\tu} \bY_u 
+ 2 \bY^\dagger_d \bm^2_{\td} \bY_d +2\bA^\dagger_u \bA_u
+2\bA^\dagger_d \bA_d \nonumber \\
&&-\frac{2}{15} M^2_1 g^2_1 -6M^2_2 g^2_2 - \frac{32}{3} M^2_3 g^2_3 .
\eeqn
The RGEs for the other soft-breaking  masses are:
\beqn{higgs}
16\pi^2 \frac{d m^2_{T} }{dt} &= & 2\left[ m^2_{T}\left(|\la_1|^2 +
{\rm Tr}( \bY^\dagger_T \bY_T)\right )+ 2 {\rm Tr}( \bY^\dagger_T \bm^2_\tl 
\bY_T) + 2 m^2_{H_1} |\la_1|^2  \right.  \nonumber \\
&& \left.+{\rm Tr}( \bA^\dagger_T \bA_T)   
+ |A_1|^2 \right] 
-\frac{24}{5} M^2_1 g^2_1 - 16 M^2_2 g^2_2  , \nonumber \\ 
16\pi^2 \frac{d m^2_{\bar{T}} }{dt} &= & 2 \left(m^2_{\bar{T}}|\la_2|^2 
+ 2 m^2_{H_2} |\la_2|^2 +|A_2|^2\right) 
-\frac{24}{5} M^2_1 g^2_1 - 16 M^2_2 g^2_2  ,\nonumber  \\ 
16\pi^2 \frac{d m^2_{S} }{dt} &= & 2 m^2_{S}
{\rm Tr}( \bY^\dagger_S \bY_S) + 4\left[ 
{\rm Tr}( \bY^\dagger_S \bm^2_\td 
\bY_S) + {\rm Tr}(\bA^\dagger_S \bA_S) \right] 
-\frac{32}{15} M^2_1 g^2_1 - \frac{80}{3} M^2_3 g^2_3  , \nonumber \\ 
16\pi^2 \frac{d m^2_{\bar{S}} }{dt} &= & 
-\frac{32}{15} M^2_1 g^2_1 - \frac{80}{3} M^2_3 g^2_3  , \nonumber \\ 
16\pi^2 \frac{d m^2_{Z} }{dt} &= & 2 m^2_{Z}
{\rm Tr}( \bY^\dagger_Z \bY_Z) + \left[ 
{\rm Tr}( \bY^\dagger_Z \bm^2_\td \bY_Z) +
{\rm Tr}( \bY^\star_Z \bm^{2T}_\tl \bY^T_Z)+ 
{\rm Tr}(\bA^\dagger_Z \bA_Z) \right] 
\nonumber \\
&& -\frac{2}{15} M^2_1 g^2_1 -{6} M^2_2 g^2_2 - 
\frac{32}{3} M^2_3 g^2_3  , \nonumber \\ 
16\pi^2 \frac{d m^2_{\bar{Z}} }{dt} &= & 
-\frac{2}{15} M^2_1 g^2_1 -{6} M^2_2 g^2_2 - 
\frac{32}{3} M^2_3 g^2_3  , \nonumber \\ 
16\pi^2 \frac{d m^2_{H_1} }{dt} &= & 
2m^2_{H_1}\left[  {\rm Tr}(\bY_e \bY^\dagger_e) + 3{\rm Tr}
(\bY_d \bY^\dagger_d) 
+ 6|\la_1|^2\right] + 
2 \left[3{\rm Tr}(\bY^\dagger_d \bm^2_\td \bY_d   
+\bY_d \bm^2_\tq \bY^\dagger_d)  \right. \nonumber \\
&& \left. +
  {\rm Tr}(\bY_e \bm^2_\tl \bY^\dagger_e +
\bY^\dagger_e \bm^2_\te \bY_e) \right] + 
6|\la_1|^2 m^2_{T}  \nonumber \\
&&+ 2\left[3{\rm Tr}(\bA_d\bA^\dagger_d)  +
{\rm Tr}(\bA_e\bA^\dagger_e) +3|A_1|^2\right] 
-  \frac{6}{5} M^2_1 g^2_1 - 6 M^2_2 g^2_2 , \nonumber \\
16\pi^2 \frac{d m^2_{H_2} }{dt} &= & 
2m^2_{H_2}\left[3{\rm Tr}
(\bY_u \bY^\dagger_u) 
+ 6|\la_2|^2\right] +  6 {\rm Tr}(\bY^\dagger_u \bm^2_\tu \bY_u    
+\bY_u \bm^2_\tq \bY^\dagger_u)   \nonumber \\
&&  +6|\la_2|^2 m^2_{\bar{T}} + 
6\left[{\rm Tr}(\bA_u\bA^\dagger_u)  + |A_2|^2\right] 
 -  \frac{6}{5} M^2_1 g^2_1 - 6 M^2_2 g^2_2 .
\eeqn
As for the soft-breaking trilinear coupling matrices we have:
\beqn{soft-A}
16\pi^2 \frac{d \bA_{T} }{dt} &= & \bA_T\left[ \bY^\dagger_e\bY_e +
9\bY^\dagger_T \bY_T +3\bY^\dagger_Z \bY_Z+ {\rm Tr}( \bY^\dagger_T \bY_T) +
|\la_1|^2 -\frac95 g^2_1-7 g^2_2\right]   \nonumber \\ 
&&+ \left( \bY^T_e \bY^\star_e + 
9 \bY_T \bY^\dagger_T +3\bY^T_Z \bY^\star_Z\right) \bA_T  + 
6  \bA^T_Z \bY^\star_Z \bY_T \nonumber \\ 
&& +  2 \bY_T\left[ 3\bY^\dagger_Z \bA_Z + 
{\rm Tr}( \bY^\dagger_T\bA_T) +\la^\star_1 A_1
+ \frac{9}{5} M_1 g^2_1  + 7 M_2 g^2_2
\right] , \nonumber \\
16\pi^2 \frac{d \bA_{S} }{dt} &= & \bA_S\left[
12 \bY^\dagger_S\bY_S +
2\bY^\star_d \bY^T_d +2\bY^\star_Z \bY^T_Z+ {\rm Tr}
( \bY^\dagger_S \bY_S) 
 -\frac45 g^2_1-12 g^2_3\right]   \nonumber \\ 
&& +2\left( 6 \bY_S\bY^\dagger_S + 
\bY_d \bY^\dagger_d +\bY_Z \bY^\dagger_Z\right) \bA_S  + 
4(  \bA_d \bY^\dagger_d+ \bA_Z \bY^\dagger_Z )\bY_S \nonumber \\ 
&& + 2 \bY_S\left[ 2\bY^\star_d \bA^T_d+2\bY^\star_Z \bA^T_Z+
{\rm Tr}( \bY^\dagger_S\bA_S) 
+ \frac{4}{5} M_1 g^2_1  + 12 M_2 g^2_3
\right] , \nonumber \\
16\pi^2 \frac{d \bA_{Z} }{dt} &= & \bA_Z\left[ \bY^\dagger_e\bY_e+
3\bY^\dagger_T\bY_T+
7 \bY^\dagger_Z\bY_Z +{\rm Tr}( \bY^\dagger_Z \bY_Z) 
 -\frac{7}{15} g^2_1-3 g^2_2 -\frac{16}{3}g^2_3\right]   \nonumber \\ 
&& +2\left( 2 \bY_S\bY^\dagger_S + 
\bY_d \bY^\dagger_d +2\bY_Z \bY^\dagger_Z\right) \bA_Z  + 
6 \bY_Z \bY^\dagger_T \bA_T \nonumber \\
&&  +2\left[ \bA_d \bY^\dagger_d+ 4\bA_S \bY^\dagger_S+ 
 \frac{7}{15} M_1 g^2_1  + 3 M_2 g^2_2 +\frac{16}{3}M_3 g^2_3
\right]\bY_Z , \nonumber \\
16\pi^2 \frac{d \bA_{e} }{dt} &= & \bA_e\left[5\bY^\dagger_e\bY_e +
3\bY^\dagger_Z\bY_Z+
{\rm Tr}( \bY_e\bY^\dagger_e +3\bY_d\bY^\dagger_d) +3|\la_1|^2 +3
\bY^\dagger_T \bY_T   -\frac95 g^2_1- 3 g^2_2 \right]\nonumber \\ 
&& +2\bY_e\left[2\bY^\dagger_e \bA_e +{\rm Tr}(\bA_e \bY^\dagger_e+3
\bA_d \bY^\dagger_d) +3 
\la_1^\star A_1 +3 \bY^\dagger_T \bA_T+ 3\bY^\dagger_Z \bA_Z 
\right.\nonumber \\
&& \left. +\frac{9}{5} M_1 g^2_1  + {3} M_2 g^2_2\right] , \nonumber \\ 
16\pi^2 \frac{d \bA_{d} }{dt} &= & \bA_d\left[5\bY^\dagger_d\bY_d + 
\bY^\dagger_u\bY_u +
{\rm Tr}( \bY_e\bY^\dagger_e +3\bY_d\bY^\dagger_d) +3|\la_1|^2 
-\frac{7}{15} g^2_1-3 g^2_2 -\frac{16}{3} g^2_3\right]
\nonumber \\ 
&& +2 \left(2\bY_S \bY^\dagger_S + \bY_Z \bY^\dagger_Z \right)\bA_d +
4\left(2\bA_S \bY^\dagger_S+\bA_Z \bY^\dagger_Z\right)\bY_d
\nonumber \\
&& +2\bY_d\left[2\bY^\dagger_d \bA_d +\bY^\dagger_u \bA_u +
{\rm Tr}(\bA_e \bY^\dagger_e+3
\bA_d \bY^\dagger_d) +3 
\la_1^\star A_1  \right. \nonumber \\
&& \left. + \frac{7}{15} M_1 g^2_1  + 
{3} M_2 g^2_2 +\frac{16}{3} M_3 g^2_3\right] , 
\nonumber \\ 
16\pi^2 \frac{d \bA_{u}}{dt} &= & \bA_u\left[5\bY^\dagger_u\bY_u + 
\bY^\dagger_d\bY_d +
3{\rm Tr}( \bY_u\bY^\dagger_u) +
3|\la_2|^2 -\frac{13}{15} g^2_1-3 g^2_2 -\frac{16}{3} g^2_3\right]
\nonumber \\ 
&& + 2\bY_u\left[2\bY^\dagger_u \bA_u  + \bY^\dagger_d \bA_d
+ 3{\rm Tr}(\bA_u \bY^\dagger_u) +3 
\la_2^\star A_2 +
\frac{13}{15} M_1 g^2_1  + {3} M_2 g^2_2 +\frac{16}{3} M_3 g^2_3\right] ,  
\nonumber \\
16\pi^2 \frac{d A_{1} }{dt} &= & A_1\left[ 
2{\rm Tr}( \bY^\dagger_e \bY_e +3\bY^\dagger_d \bY_d) +{\rm Tr}
(\bY^\dagger_T \bY_T) +
21 |\la_1|^2 -\frac95 g^2_1-7 g^2_2\right]   \nonumber \\
&& 
+ 2 \la_1\left[ {\rm Tr}( \bY^\dagger_T\bA_T )
+ 2{\rm Tr}( \bY^\dagger_e\bA_e +3 \bY^\dagger_d\bA_d)
 +\frac{9}{5} M_1 g^2_1  + 7 M_2 g^2_2
\right] , \nonumber \\
16\pi^2 \frac{d A_{2} }{dt} &= & A_2\left[ 
6{\rm Tr}(\bY^\dagger_u \bY_u) 
+21 |\la_2|^2 -\frac95 g^2_1-7 g^2_2\right]  \nonumber \\
&& +2 \la_2\left[ 6{\rm Tr}( \bY^\dagger_u\bA_u )
 +\frac{9}{5} M_1 g^2_1  + 7 M_2 g^2_2\right] .
\eeqn
Clearly, all the RGEs shown above are valid in both  scenarios (A) and 
(B). In the former scenario, the condition  $\bY_S= \bY_Z=0$ at $M_G$ 
ensures that Yukawa couplings $\bY_S, \bY_Z$ and the parameters 
$\bA_S, \bA_Z$ are not radiatively induced, 
therefore they can be simply switched off in the r.h.s of any 
RGEs. 
Beneath the mass scale $M_T$, we recover the RGEs of the MSSM 
by switching off $\bY_{T,S,Z}, \la_{1,2}$, 
$m^2_{T ,\bar{T}},m^2_{S ,\bar{S}} ,m^2_{Z ,\bar{Z}}$ and $\bA_{T,S,Z}, 
A_{1,2}$.  
For completeness, we report also the RGE of the $d=5$ neutrino-operator 
Yukawa matrix $\bY_\nu$ valid below $M_T$ 
in the MSSM \cite{RG-nu}:
\be{rge5} 
16\pi^2 \frac{d \bY_\nu}{dt} =\bY_\nu \left[
 -\frac{6}{5} g^2_1 - 6 g^2_2  + 6{\rm Tr}(\bY^\dagger_u \bY_u)
\right] + 
 \bY_\nu \bY^\dagger_e \bY_e + (\bY^\dagger_e \bY_e)^T \bY_\nu .
\ee

\end{document}